# *SatTrack*: Software for Evaluating Satellite Interference and Rim-Based Interference Mitigation Using a Reconfigurable Parabolic Antenna

J. M. Santana, L. Heller, and R.M. Buehrer
Virginia Polytechnic Institute and State University
Blacksburg, VA
justins04@vt.edu, lukeh28@vt.edu

*Abstract*—Large Low Earth Orbit (LEO) constellations (e.g., Starlink and Iridium) significantly increase the likelihood of transient, high-power interference events at ground receivers. This report presents *SatTrack*, a GUI-driven simulation framework that (i) tracks satellite motion relative to a fixed antenna boresight, (ii) predicts reflector gain patterns of a parabolic reflector antenna with a reconfigurable rim of specified size using a physical optics (PO) surface-current model decomposed into fixed and reconfigurable rim regions, and (iii) synthesizes deep, directional nulls using fast rim-weight optimization algorithms. Beyond baseline serial greedy and greedy bit-flip methods, additional files also support advanced weight optimization algorithms, including simulated annealing and majorization-minimization optimizers operating over higher-order complex weight alphabets, enabling deeper null formation. Interference power from satellites is estimated using free-space path loss and an FCC-derived Starlink EIRP versus steering-angle model, and event-level throughput for each satellite is computed using Shannon capacity with a throughput cap. Simulation results demonstrate that while simple binary rim control achieves about 40-55dB of average interference suppression, advanced optimization methods can exceed 65-70dB of average suppression under favorable geometries, with best-case received interference power below -220dBm. We compare three scenarios: No interference mitigation, muting satellites during interference events, and rim-based spatial nulling. We show that rim-based nulling can achieve nearly the same interference suppression as muting satellites while allowing nearly the same satellite throughput as no mitigation. These results highlight both the scalability of rim-based reflector reconfiguration and the fundamental limitation imposed by satellites crossing directly through the main lobe.

I. INTRODUCTION

Low Earth Orbit (LEO) satellites traverse the sky at high angular velocities, producing short-duration but potentially severe "in-beam" interference events for high-gain parabolic antennas. As LEO constellations continue to grow in size and density, these transient interference events become increasingly frequent, motivating the need for adaptive mitigation strategies that operate on sub-second time scales without requiring full mechanical repointing of the antenna.

A practical mitigation approach is to maintain a fixed main reflector while electronically reconfiguring only a narrow rim region to reshape sidelobes and introduce deep nulls in the directions of interfering satellites. Prior work has shown that the far-field radiation pattern of a reflector can be decomposed into fixed and reconfigurable contributions, with rim currents providing sufficient degrees of freedom to achieve effective pattern control [1,2]. This decomposition enables fast, low-complexity null synthesis while preserving the main-beam performance of the dish.

In this work, the rim-scattering concept is implemented within SatTrack, a simulation framework that couples satellite tracking, physical-optics-based antenna modeling, and per-event null optimization. The baseline implementation employs a binary (±1) rim-weight model optimized using serial greedy and greedy bit-flip algorithms, enabling rapid interference suppression suitable for real-time operation. These methods already provide substantial mitigation, reducing received interference power by tens of decibels across a wide range of observation geometries.

To further explore the performance limits of rim-based interference mitigation, this report also presents advanced optimization techniques that operate over higher-order complex weight alphabets. Section X extends the baseline framework by incorporating simulated annealing and majorization-minimization (MM) methods, allowing continuous-phase control on the rim currents and significantly deeper null formation. These advanced methods take inspiration from previous works [3, 4]. These advanced methods are evaluated across multiple boresight angles and satellite densities, demonstrating average suppression levels exceeding those of binary control, while also highlighting diminishing returns in dense or main-lobe-dominated scenarios.

Together, the baseline and advanced results illustrate the trade space between algorithmic complexity, achievable null depth, and practical limitations imposed by antenna geometry and satellite trajectories. The combined findings underscore both the effectiveness and the fundamental constraints of reconfigurable rim-based interference mitigation for modern LEO environments.

## II. SATTRACK SYSTEM OVERVIEW

SatTrack provides an interactive workflow to:

- Detect Starlink/Iridium satellites crossing the beam,
- Compute fixed, rim, and total gain patterns,
- Compute null depth and interference power per event,
- Estimate throughput and seconds in beam per event.

The GUI is run with a local frontend and a Python backend server. Users select tracking mode (manual or celestial body), satellite constellation, ground location, time window, sampling interval, and antenna parameters. A key simulation assumption is that each detected satellite event is treated independently (when nulling one satellite, the model assumes no other satellites are simultaneously in beam). Note that the algorithms are capable of nulling multiple signals at once [2,3,4], but that is not currently included in this software.

## III. SATELLITE EVENT GEOMETRY

For each detected event, the backend computes an "offset direction" in the antenna's local pointing coordinates. Specifically, it uses the event's azimuth and altitude offsets $\Delta Az$ and $\Delta Alt$ and defines:

$$\theta_{deg} = \sqrt{(\Delta Az)^2 + (\Delta Alt)^2}, \qquad \phi = \tan^{-1}\left(\frac{\Delta Alt}{\Delta Az}\right)$$

then converts $\theta$ to radians for the antenna model. The pointing angle $(\theta, \phi)$ is the target null direction used by the rim weight optimization algorithm for that event.

## IV. ANTENNA SIMULATION MODEL

SatTrack reflector simulation is a PO style sum over discrete surface patches. The implementation builds:

1. A patch grid on the paraboloid,
2. An incident feed magnetic field $H_i$ at each patch,
3. An equivalent surface current $J_0$,
4. A far field sum with a phase term $e^{jk\hat{r}\cdot r_p}$, split into fixed vs rim contributions, then projected into a co-polar basis.

### A. Reflector Patch Grid on the paraboliod antenna surface

The reflector surface is discretized into rings of patches. Two arc length spacings are used:
- Inner region step $dl_0 = dlwl0 \cdot \lambda$
- Outer region step $dl_1 = dlwl1 \cdot \lambda$

For a ring radius $\rho$, angular step is:

$$d\phi = \frac{dl}{\rho}, \qquad \phi \in [d\phi, 2\pi]$$

And the paraboloid surface uses:

$$z(\rho) = \frac{\rho^2}{4f}$$

Each patch has an associated surface area weight:

$$\Delta s(\rho) = \frac{dl^2 \sqrt{4f^2 + \rho^2}}{2f}$$

The reconfigurable rim mask is:

$$\rho > \frac{D_0}{2}$$

The fixed region is:

$$\rho \leq \frac{D_0}{2}$$

### B. Feed Magnetic Field Model $H_i(r_p)$

Let the feed be located at:

$$P_f = [0,0,f]^T$$

And for each patch position $r_p$, define:

$$s_i = r_p - P_f, \qquad R = \|s_i\|, \qquad \hat{s}_i = \frac{s_i}{R}$$

The feed polarization reference vector is:

$$\hat{l}_{eh} = [0,1,0]^T$$

where:

$$f = \hat{l}_{eh} \times \hat{s}_i, \qquad \hat{u}_{pol} = \frac{f}{\|f\|}$$

The spreading/phase term and taper are:

$$phase\_spread = \frac{e^{-jkR}}{R}, \qquad taper = (-\hat{s}_{i,z})^q$$

Thus, the incident magnetic field is:

$$H_i(r_p) = H_0 \hat{u}_{pol} \frac{e^{-jkR}}{R} (-\hat{s}_{i,z})^q$$

with $H_0 = 10^{-3}$.

### C. Patch Normal and PO Equivalent Surface Current

The code constructs a patch normal $\hat{n}$ using intermediate vectors:

- $r_f = r_p - P_f$, $\hat{r}_f = \frac{r_f}{\|r_f\|}$,
- $\rho = \sqrt{x^2 + y^2}$,
- $\theta_f = \tan^{-1}\left(\frac{\rho}{f-z}\right)$,
- And tangent vector:

$$t_f = \begin{bmatrix} \cos\phi \cos\theta_f \\ \sin\phi \cos\theta_f \\ \sin\theta_f \end{bmatrix}$$

Then:

$$\hat{n} = -\hat{r}_f \cos\left(\frac{\theta_f}{2}\right) + t_f \sin\left(\frac{\theta_f}{2}\right)$$

The PO equivalent surface current is:

$$J_0 = 2(\hat{n} \times H_i)$$

And the per element contribution stored is:

$$q_{elem} = J_0 \Delta s$$

D. Far Field Sum, Transverse Projection, and Co-polar Component

For a look direction $(\theta, \phi)$, the code forms the unit vector:

$$\hat{r}(\theta, \phi) = \begin{bmatrix} \sin\theta \cos\phi \\ \sin\theta \sin\phi \\ \cos\theta \end{bmatrix}$$

And applies a phase factor per patch:

$$e^{jk\hat{r}\cdot r_p}$$

to form $q_{all} \odot e^{jk\hat{r}\cdot r_p}$.

The fixed and rim sums are computed by summing $q_{all}$ over the fixed mask and rim mask, with rim patched multiplied by binary weights $w \in \{-1, +1\}$.

Before converting to an electric field, each summed vector is projected transverse to $\hat{r}$:

$$q_t = q - (q \cdot \hat{r})\hat{r}$$

Next, the co-polar basis is defined by:

$$\hat{a}_{az} = \begin{bmatrix} \cos\phi \\ -\sin\phi \\ 0 \end{bmatrix}, \quad \hat{a}_{el} = \begin{bmatrix} \cos\theta \sin\phi \\ \cos\theta \cos\phi \\ -\sin\theta \end{bmatrix}$$

Then:

$$v = (\hat{a}_{az} \times \hat{r}) + j(\hat{r} + \hat{a}_{el}), \quad \hat{e}_{co} = \frac{v}{\|v\|}$$

Finally, the co-polar electric field contributions are computed with:

$$E = \frac{-j2\pi f \mu_0}{4\pi} \cdot (q_t \cdot \hat{e}_{co})$$

## V. RIM NULLING OPTIMIZATION

### A. What The Optimizer is Minimizing

At a target direction $(\theta, \phi)$, the code reduces the problem to a complex scalar cancellation in the co-polar component. It builds per-rim element scalars (after transverse projection + co-pol projection):

$$q_i \in \mathbb{C}, \quad i = 1, \ldots, N$$

And a fixed region scalar $q_{fixed}$.

With binary weights $w_i \in \{-1, +1\}$, the co-pol sum is:

$$S(w) = q_{fixed} + \sum_{i=1}^{N} w_i q_i$$

and the objective is to minimize $|S(w)|$ in the target direction.

### B. Serial Greedy Search (One Pass)

The serial method iterates $i = 1 \ldots N$, and at each step chooses $w_i$ to reduce the magnitude:

$$w_i = \begin{cases} +1, & |S + q_i| < |S - q_i| \\ -1, & otherwise \end{cases}$$

### C. Greedy Bit-Flip

A second weight vector starts from all $+1$ and tries flipping each element once. Flipping $w_i$ changes the sum by $-2q_i$:

$$S_{try} = S_{curr} = -2q_i$$

And the flip is kept if it improves $|S|$.

### D. Mapping "bit" Encoding to $\pm 1$ Weights

The implementation stores $w \in \{0, 1\}$ then maps to $\pm 1$ via:

$$w_i = 2 bit_i - 1$$

## VI. GAIN OUTPUTS AND NORMILIZATION

For a scan over $\theta$ values, SatTrack reports:

- $G_0$: fixed only gain,
- $G_1$: rim only gain (serial),
- $G_t$: total gain (serial),
- $G_{1b}, G_{tb}$: rim/total gains (bit-flip).

Co-pol power gain from the co-pol electric field is computed for each component using:

$$G = \frac{|E_{co}|^2}{\eta}$$

This gain is then divided by a constant normalization factor:

$$\frac{7.2161 \cdot 10^{-4}}{4\pi}$$

## VII. INTERFERENCE AND THROUGHPUT MODELS

### A. Interference "Received Power"

Received power is calculated from EIRP and antenna gain with the free space path loss equation:

$$FSPL = 20 \log_{10}\left(\frac{4\pi df}{c}\right), \qquad P_r = EIRP + G - FSPL$$

This interference power is plotted for fixed/serial/bit-flip gains on the SatTrack GUI.

### B. Event EIRP Model Used for Throughput

*EIRP* density vs. steering angle for Starlink satellites is defined by the graph below [5]:

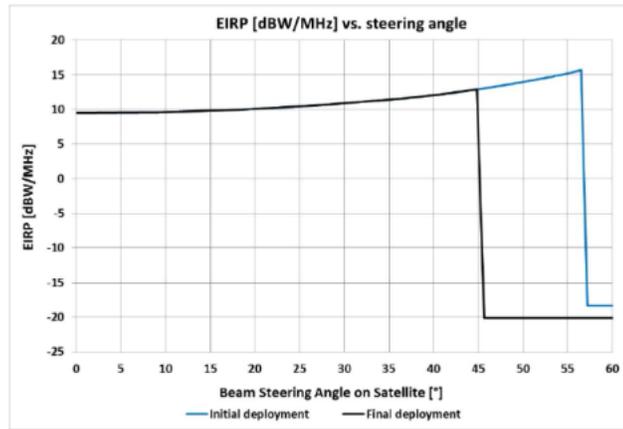

Figure A.3.2-1: EIRP Density Variation by Beam Steering Angle

*Figure 1: EIRP density vs. steering angle for Starlink satellites per FCC documentation.*

Using interpolation, this plot is recreated in the SatTrack codebase. For interference and throughput simulations, SatTrack uses this function below:

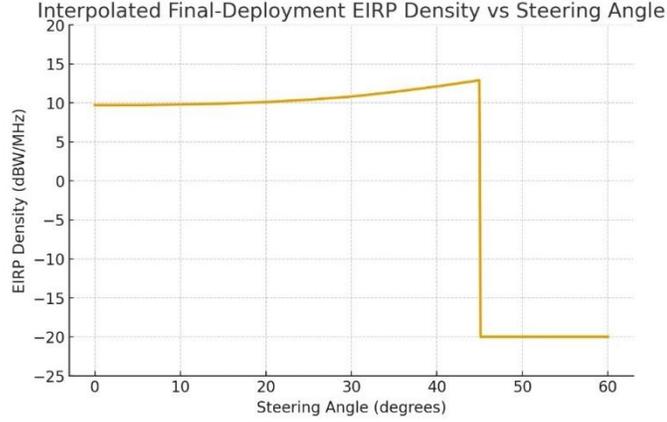

*Figure 2: Interpolated EIRP density vs. steering angle for Starlink satellites.*

Total *EIRP* over bandwidth $B$ is then calculated:

$$EIRP = density + 10\log_{10}(B)$$

*C. Throughput Calculation*

For each event $i$, SatTrack uses:
- Bandwidth $B$ ($Hz$),
- Event slant range $d_i$,
- Event $EIRP_i$ ($dBm$)
- Receive gain $G_{rx,i}$ ($dBi$),
- Noise figure $NF$ ($dB$),
- Misc losses $L$ ($dB$),
- Per-event dwell time $dt$.

Step 1: Thermal noise power ($dBm$)

$$N = -174 + 10\log_{10}(B) + NF$$

Step 2: Free space path loss using $f = 11.7\ Ghz$

$$FSPL_i = 20\log_{10}\left(\frac{4\pi d_i}{c/11.7\ GHz}\right)$$

Step 3: Received power ($dBm$)

$$P_{r,i} = EIRP_i + G_{rx,i} - FSPL_i - L$$

Step 4: SNR and linear conversion

$$SNR_i = P_{r,i} - N, \qquad SNR_{i,lin} = 10^{\frac{SNR_i}{10}}$$

Step 5: Shannon capacity/rate ($bps$) with spectral efficiency cap $\eta_{max}$

$$R_i = B\log_2(1 + SNR_{i,lin}), \qquad R_i = \min(R_i, B\eta_{max})$$

Step 6: Bits accumulated for the event

$$bits_i = R_i(duty\_cycle)$$

The SatTrack system assumes the receive gain is $34\ dBi$, which is the gain of the Starlink user terminal. The bandwidth is set to $240\ MHz$. All simulations are also run at $11.7\ GHz$, because Starlink transmits between $10.7 - 12.7\ GHz$. All these parameters are listed in a Starlink FCC application [5].

## VIII. USING THE GUI

SatTrack is designed to provide an interactive, end-to-end workflow for simulating satellite interference events and evaluating rim-based mitigation strategies. The graphical user interface (GUI) serves as the primary user interaction layer, allowing all simulation parameters to be configured without modifying source code.

### A. Launching the GUI

The SatTrack GUI is run locally and consists of a web-based frontend and a Python backend server. To launch the system, two terminal windows are required. In the first terminal, the frontend is started by activating the Python virtual environment, installing dependencies, and running the development server. In the second terminal, the backend API is launched using the same virtual environment and required Python packages. Once both processes are running, the GUI can be accessed through a web browser at http://localhost:5173/.

### B. Observation and Tracking Configuration

The left sidebar of the GUI contains all configurable simulation parameters. The **Tracking Mode** determines how the antenna boresight is defined. In *Manual* mode, the antenna is pointed to a user specified azimuth and elevation. In *Celestial Body* mode, the antenna automatically tracks a selected astronomical object (e.g., Polaris), allowing simulations to reflect realistic fixed-point observations.

Users select the satellite **Constellation** (Starlink or Iridium) and the **Observation Location**, which defines the antenna's geographic position. The **Simulation Time** parameters specify the total duration (hours and minutes) and the **Sampling Interval**, which controls how often satellite positions are evaluated. These parameters directly influence the number of detected satellite events and the temporal resolution of the results.

### C. Antenna and Nulling Parameters

The **Antenna** section allows the user to define the physical and electrical characteristics of the reflector, including total diameter, reconfigurable rim width, beamwidth, operating frequency, feed taper exponent, and angular resolution. Optional Target Theta and Target Phi inputs allow users to manually generate and visualize antenna gain patterns for a specific null direction without running a full satellite simulation.

The **Nulling Methods** section enables or disables different mitigation strategies on the output plots. Users can independently toggle fixed-gain performance, serial greedy nulling, and greedy bit-flip nulling. These controls allow direct visual comparison between unmitigated and mitigated cases within the same simulation run.

### D. Running a Simulation

To execute a simulation, all required fields must be populated except for the optional target angles and satellite filter. After clicking **Run Simulation**, the backend computes satellite detection events, antenna gain patterns, null depths, interference power, and throughput metrics. Depending on the simulation duration and satellite density, the computation may take one to several minutes. Once complete, all plots are automatically populated in the main display area.

Each detected satellite crossing is treated as an independent event. When nulling is applied, the simulation assumes that only a single satellite is present in the beam at a given time.

### E. Visualization and Data Interpretation

The GUI presents multiple synchronized plots, including antenna null depth, received interference power, detected event geometry, and throughput verses time. Hovering over any plot reveals detailed per-event information such as scan angle, null depth, received power, and detected event time. The **Satellite Filter** dropdown can be used to isolate a single satellite, reducing visual clutter in dense simulations such as wide beam Starlink scenarios.

An additional **Antenna** view allows users to visualize the simulated gain pattern directly. Fixed, rim-only, and total gain contributions are displayed, and legend entries can be toggled to show or hide individual components. The reflector geometry is drawn to scale, with the reconfigurable rim clearly indicated, providing intuitive insight into how rim weighting affects the far field pattern.

## IX. EXAMPLE RESULTS

All simulation results in this section used the following parameters:
- $D_{total} = 2.5m$
- $D_{rim} = 0.278m$
- $f = 0.972m$
- $f_c = 11.7 Ghz$
- $resolution = 0.05°$

The location was set to Blacksburg with the dish pointing at an altitude of 40° and an azimuth of 0°. The simulation tracked the Starlink constellation, with an observation time of 5 minutes at a sampling rate set to once per 5 seconds. Each detected satellite is represented by a different colored dot on each plot. The goal was to compare the impact of satellite transmissions on a radio astronomy collection in three different scenarios: (a) No mitigation; (b) Muting Satellites; and (c) Spatial Nulling using the reconfigurable rim. In the first scenario (no mitigation), any data collected with interference must be discarded, but there is no impact on satellite throughput. In the second scenario, satellites are muted whenever they are within a predefined range of angles relative to boresight. This removes interference during those times but degrades satellite throughput. In the third scenario, interference mitigation is used such that ideally, very little satellite throughput is lost, and very little collected data (ideally none) is discarded. Note that in this case, the antenna is not capable of nulling interference when a satellite is passing through the main beam. In those cases, either data could be discarded or the satellite could be muted. However, due to the narrow beamwidth (roughly $1° - 2°$), such events are rare, as we will see. For each scenario, we considered three ranges of angles relative to boresight to define interference events. The main beamwidth is $1° - 2°$. Thus, at roughly 5° from boresight, the gain is very low relative to boresight. However, to be thorough, we also considered interference events within 15° of boresight and within 45° of boresight.

### A. Considering only events within 5° of boresight

**GUI Results:**

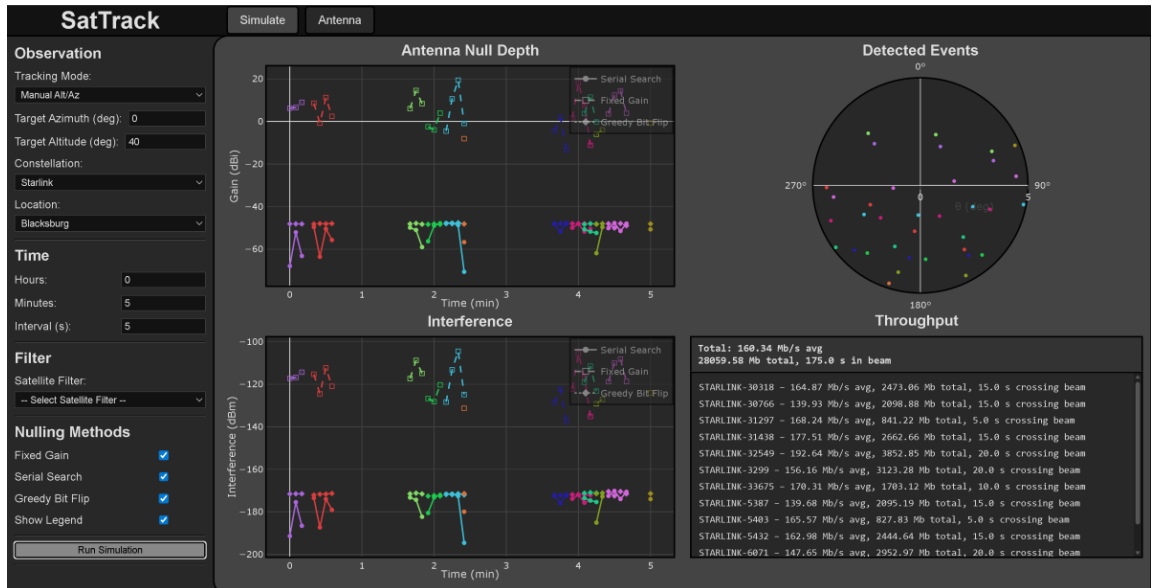

*Figure 3: Simulation results tracking Starlink events within 5° boresight over 5 minutes.*

As shown in figure 3, the interference mitigation techniques presented in this paper drastically reduce the received interference power over all detected events.

**Interference Power (no mitigation):**

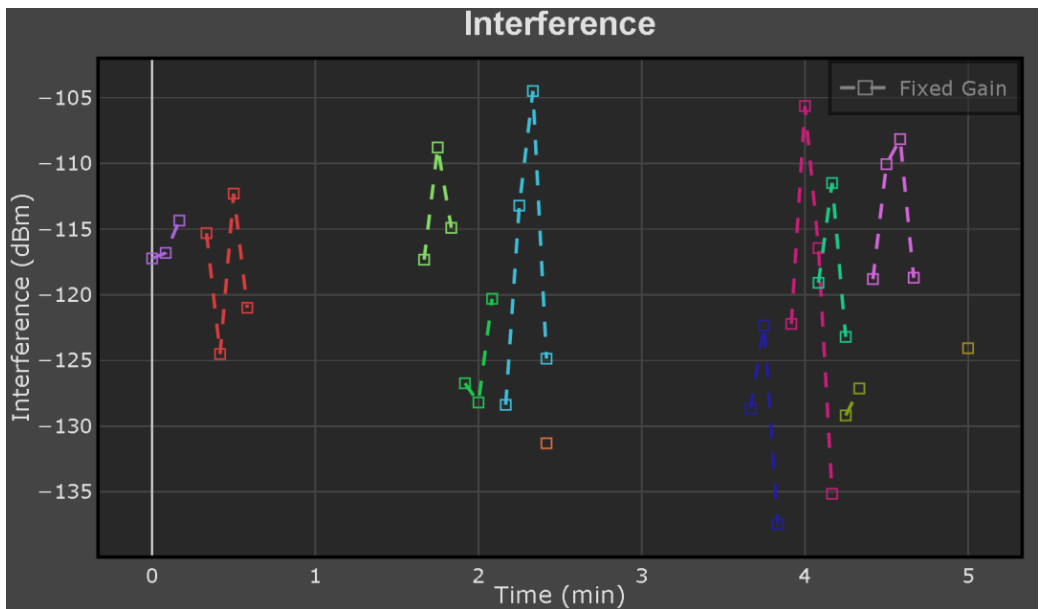

*Figure 4: Interference power for detected events with no mitigation tracking Starlink events within 5° boresight over 5 minutes.*

Based on figure 4, the average interference power with no mitigation for Starlink satellites crossing a 5° boresight over 5 minutes is $\approx -117 dBm$. The highest interference power is $-104.49 dBm$, and the lowest interference power is $-137.47 dBm$.

**Interference Power (serial, greedy bit-flip):**

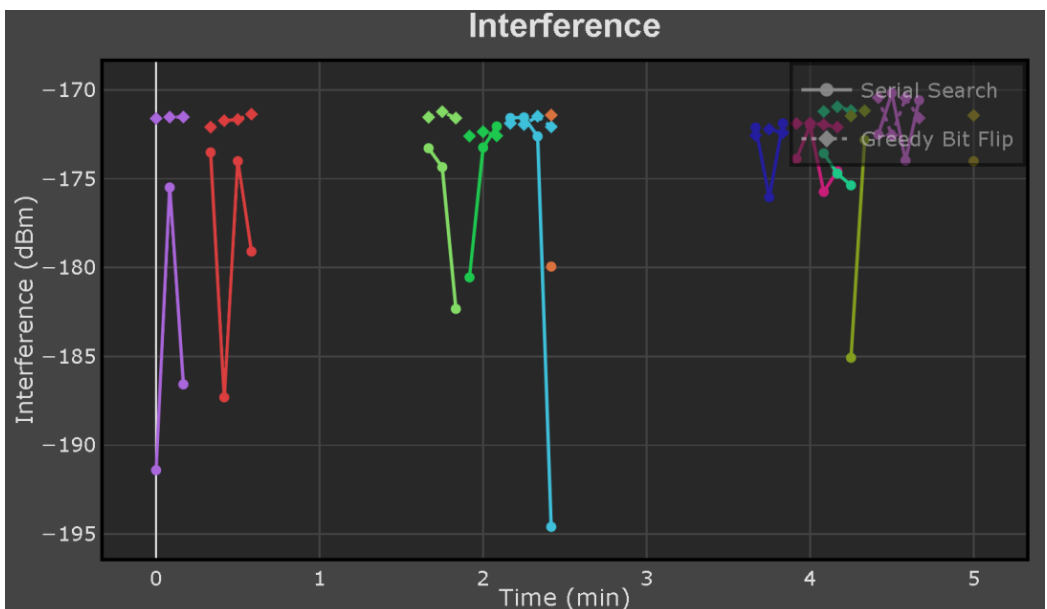

*Figure 5: Interference power for detected events using serial search/greedy bit-flip mitigation tracking Starlink events within 5º boresight over 5 minutes.*

Based on figure 5, both serial search and greedy bit-flip decrease the interference power by approximately 55 $dB$. Serial search slightly outperforms greedy bit-flip in terms of null depth.

**Throughput Lost If All Satellites Are Disabled While Crossing the Boresight:**

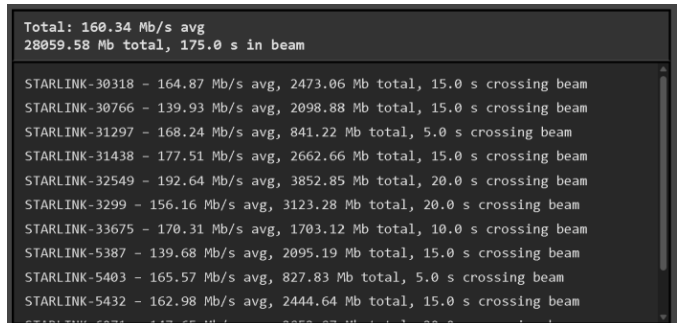

*Figure 6: Total average throughput of detected Starlink satellites crossing a boresight of* 5°.

Based on figure 6, the total average throughput of Starlink satellites crossing a boresight of 5° is 160.34 $Mb/s$, with a total of 28059.58 $Mb$ transmitted over a total crossing time of 175 seconds.

*B. Considering only events within* 15° *of boresight*

**GUI Results:**

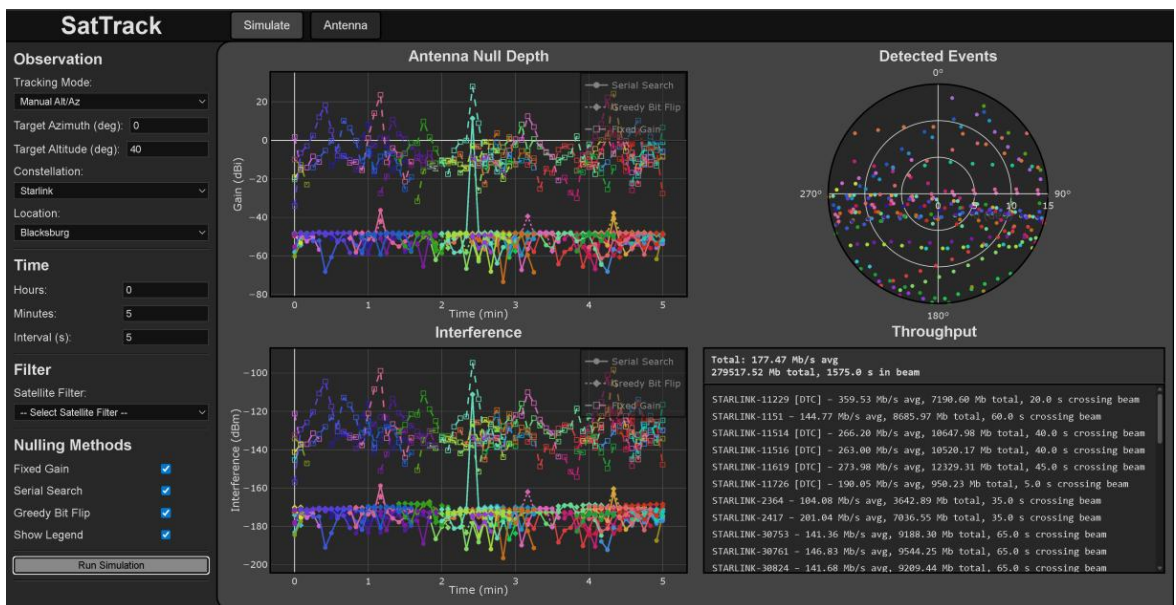

*Figure 7: Simulation results tracking Starlink within 15° of boresight over 5 minutes.*

As shown in figure 7, the interference mitigation techniques presented in this paper drastically improve the received interference power over all detected events.

**Interference Power (no mitigation):**

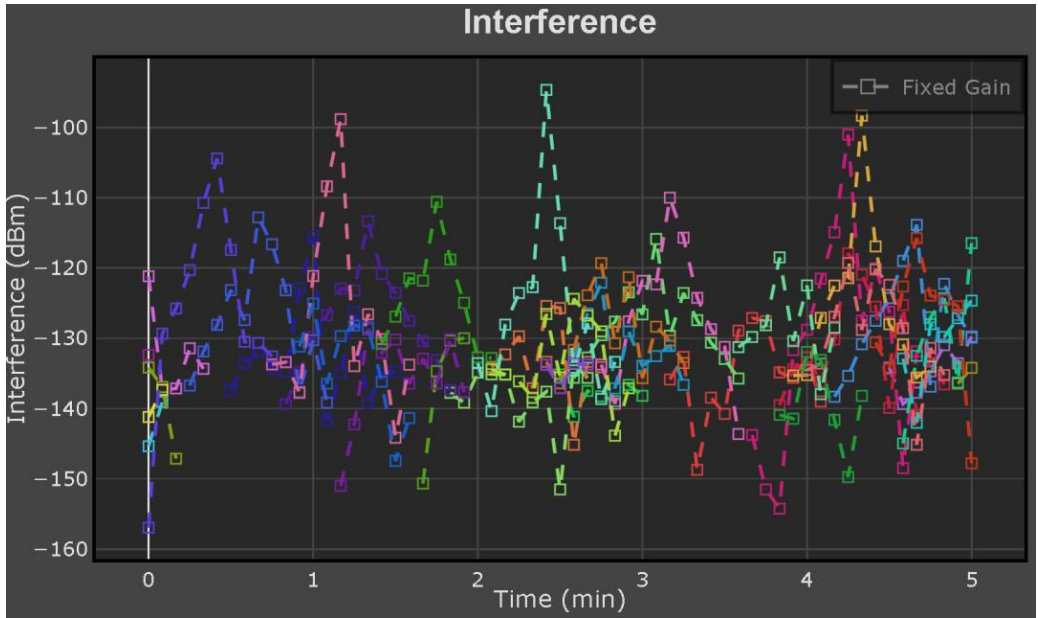

*Figure 8: Interference power for detected events with no mitigation tracking Starlink events within $15°$ boresight over 5 minutes.*

Based on figure 8, the average interference power with no mitigation for Starlink satellites crossing a 15° boresight over 5 minutes is $\approx -130 dBm$. The highest interference power is $-94.64 dBm$, and the lowest interference power is $-156.95 dBm$.

**Interference Power (serial, greedy bit-flip):**

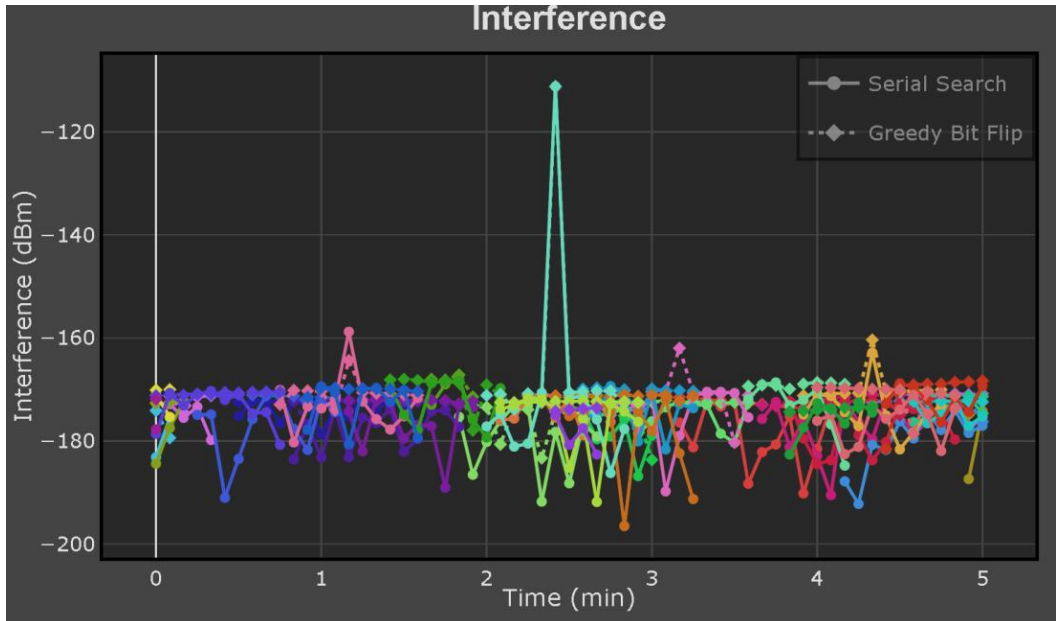

*Figure 9: Interference power for detected events using serial search/greedy bit-flip mitigation Starlink events within $15°$ of boresight over a 5-minute period.*

Based on figure 9, both serial search and greedy bit-flip decrease the interference power by $\approx -40\ dB$. Serial search slightly outperforms greedy bit-flip in terms of null depth. The large spike in the middle is caused by a satellite crossing the main lobe.

**Throughput Lost If All Satellites Are Disabled While Crossing the beam:**

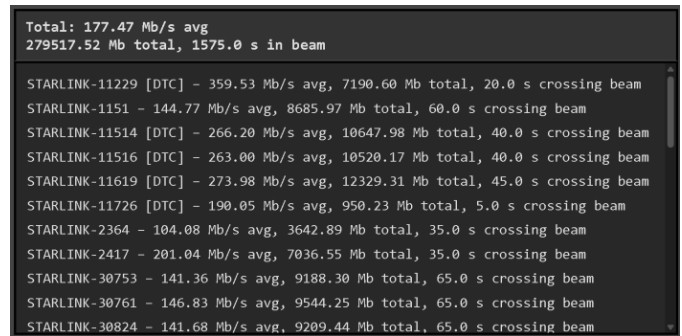

*Figure 10: Total average throughput lost due to muting Starlink satellites crossing within* 15° *of boresight.*

Based on figure 10, the average lost throughput of Starlink satellites crossing a boresight of 15° is **177.5Mbps** with a total lost throughput of 279517.52 *Mb* transmitted over a total crossing time of 1575 seconds.

*C. Considering only events within 45º of boresight*

**GUI Results:**

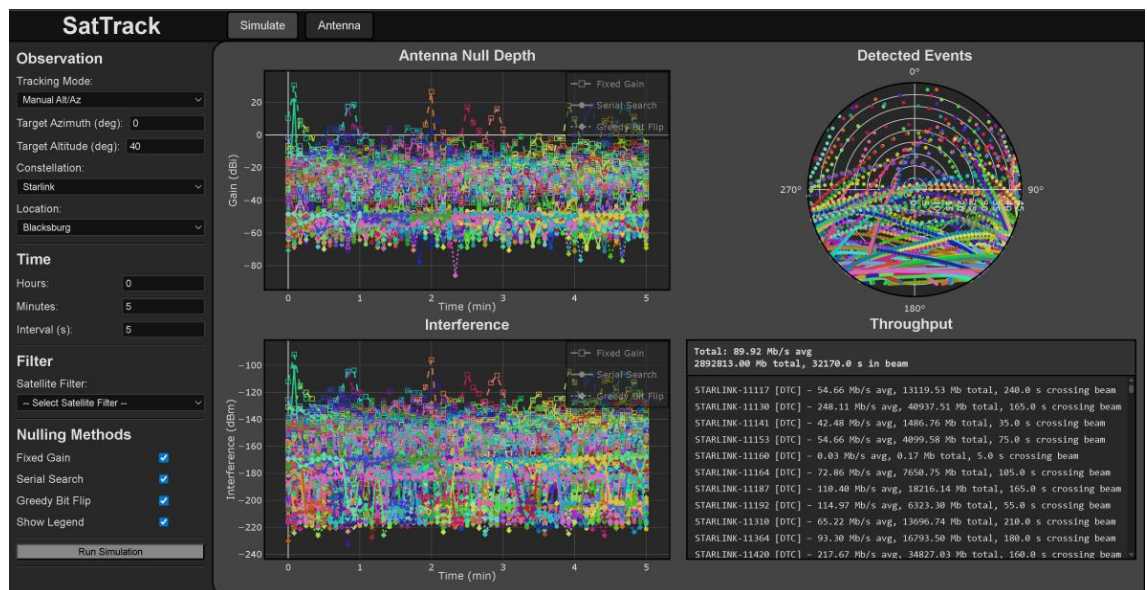

*Figure 11: Simulation results tracking Starlink within 45° of boresight over 5 minutes.*

At the time of the simulation, there are 9,171 Starlink satellites in orbit, resulting in 6,913 detected events. This high number of events clutters the graph and makes the results difficult to interpret. The Detected Events graph appears especially congested near the bottom because the field of view encompasses an increasing number of orbital paths as the altitude approaches 0°. With the boresight set to 45° and the observation angle set to 40°, the simulation includes all orbital paths down to the horizon. This is reflected in the sharp cutoff visible at the bottom of the graph.

**Interference Power (no mitigation):**

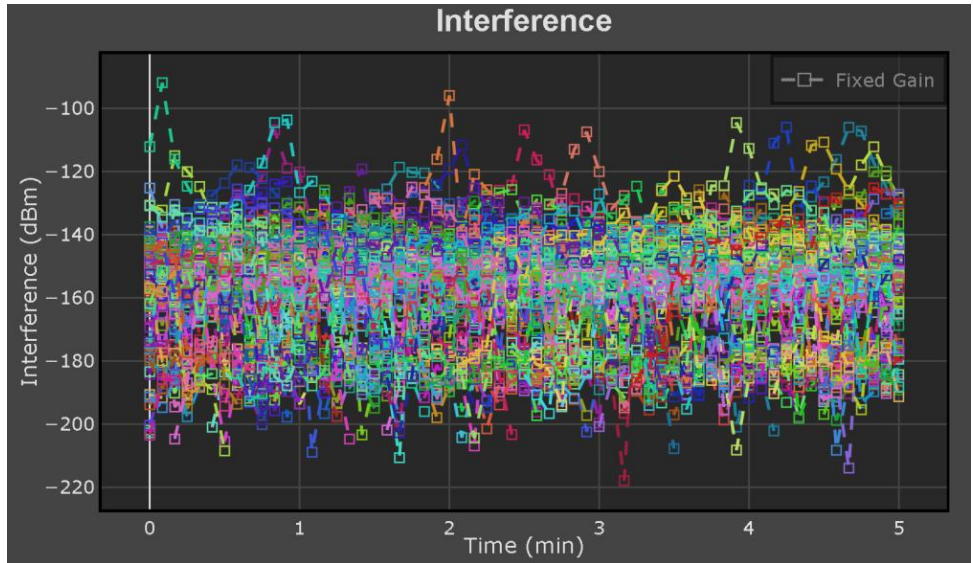

*Figure 12: Interference power for detected events with no mitigation over a 5-minute Starlink simulation with a 45° boresight.*

Based on figure 12, the average interference power with no mitigation for Starlink satellites crossing a 45° boresight over 5 minutes is $\approx -160 dBm$. The highest interference power is $-91.87 dBm$, and the lowest interference power is $-217.89 dBm$.

**Interference Power (serial, greedy bit-flip):**

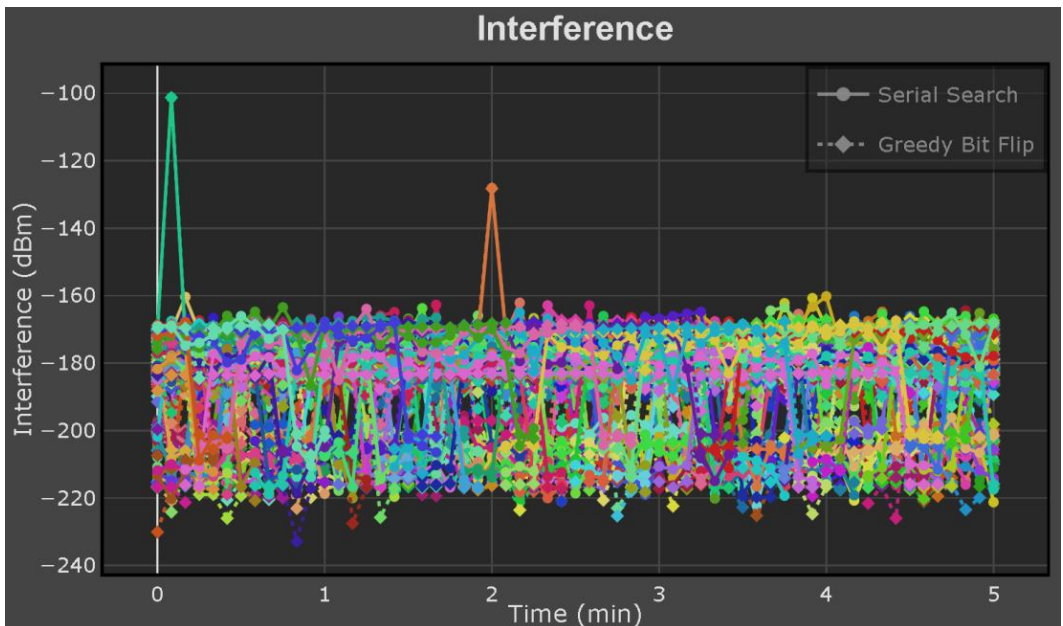

*Figure 13: Interference power for detected events using serial search/greedy bit-flip mitigation, over a 5-minute Starlink simulation with a 45° boresight.*

Based on figure 13, both serial search and greedy bit-flip decrease the interference power by $\approx -38\ dB$. Due to the large number of satellites in the graph, and the large boresight, it is difficult to tell if serial search or greedy bit flip works better. It seems as if they both have the same average null strength across all 45°.

**Throughput Lost If All Satellites Are Disabled While Crossing the Boresight:**

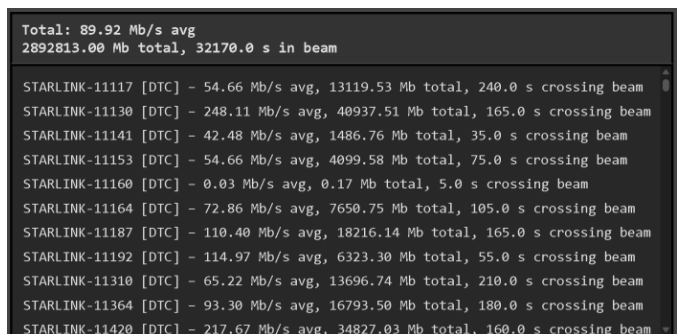

*Figure 14: Total average throughput of detected Starlink satellites crossing within 45° of boresight.*

Based on figure 14, the total average throughput of Starlink satellites crossing a boresight of 45° is 89.92 $Mb/s$, with a total of 2892813.00 $Mb$ transmitted over a total crossing time of 32170.0 seconds.

D. Results Using Serial Search Interference Mitigation
**5° from Boresight:**

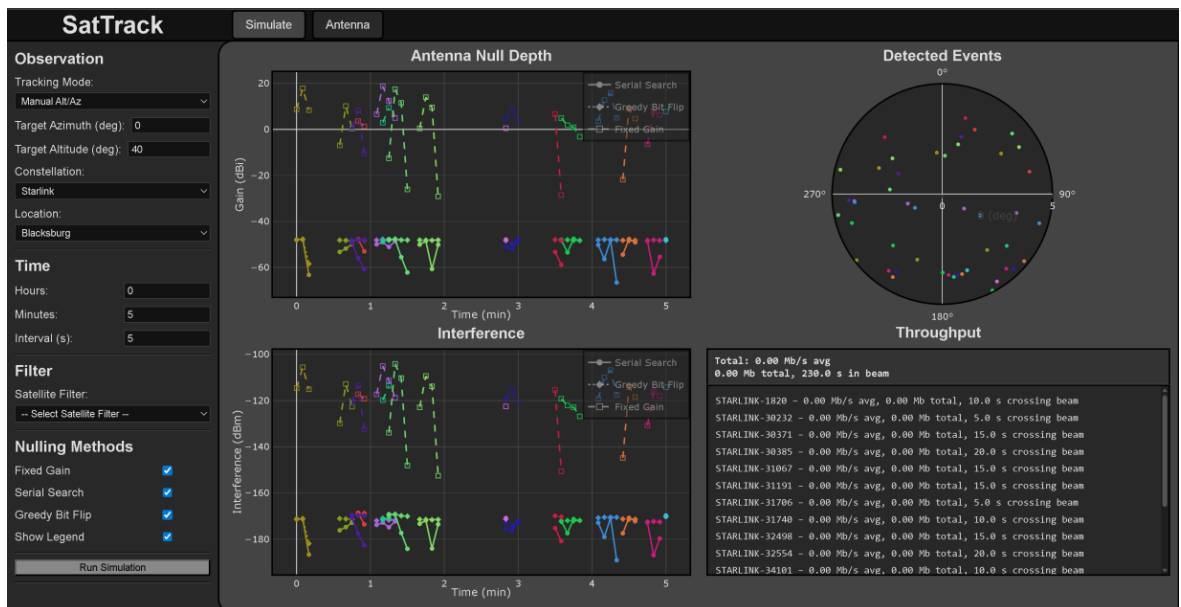

*Figure 14: Simulation results using serial search interference mitigation gain pattern for throughput calculations for a 5-minute Starlink simulation with a boresight of 5°.*

**15° from Boresight:**

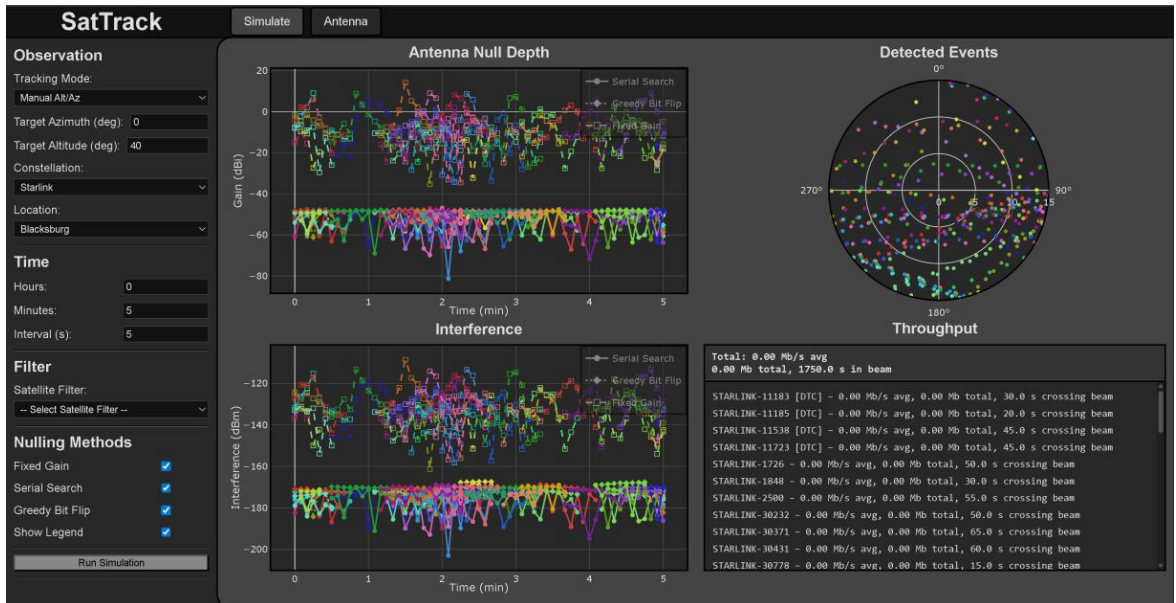

*Figure 15: Simulation results using serial search interference mitigation gain pattern for throughput calculations for a 5-minute Starlink simulation with a boresight of 15°.*

**45° from Boresight:**

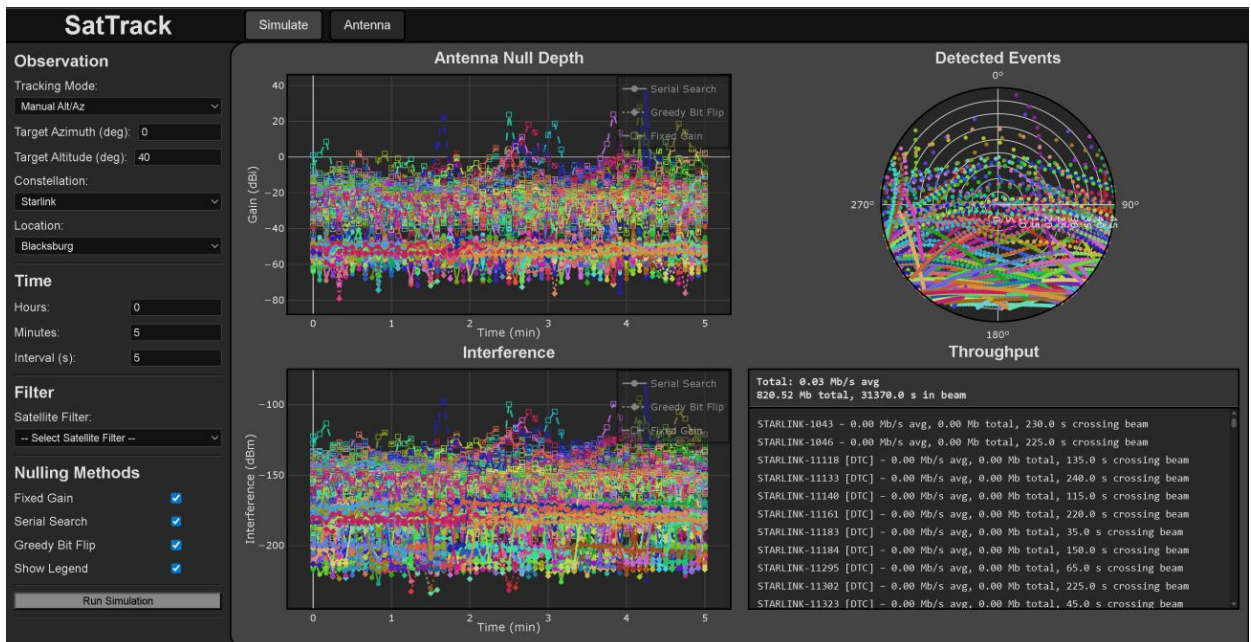

*Figure 16: Simulation results using serial search interference mitigation gain pattern for throughput calculations for a 5-minute Starlink simulation within 45° of boresight.*

As the simulation results show, using serial search to null interference results in an average received throughput of ~0 *Mb* on the simulated parabolic antenna. The only received throughput is due to satellites passing directly through the main lobe. This throughput is negligible since there are 6,913 detected events, and only 2 satellites crossed directly through the main lobe. This results in a 0.0289% chance of a Starlink satellite causing noticeable interference over a 5-minute simulation.

The small amount of throughput with a boresight of 45° is caused by STARLINK-3185 and STARLINK-6094 crossing directly through the main lobe, as shown in figures 17 & 18.

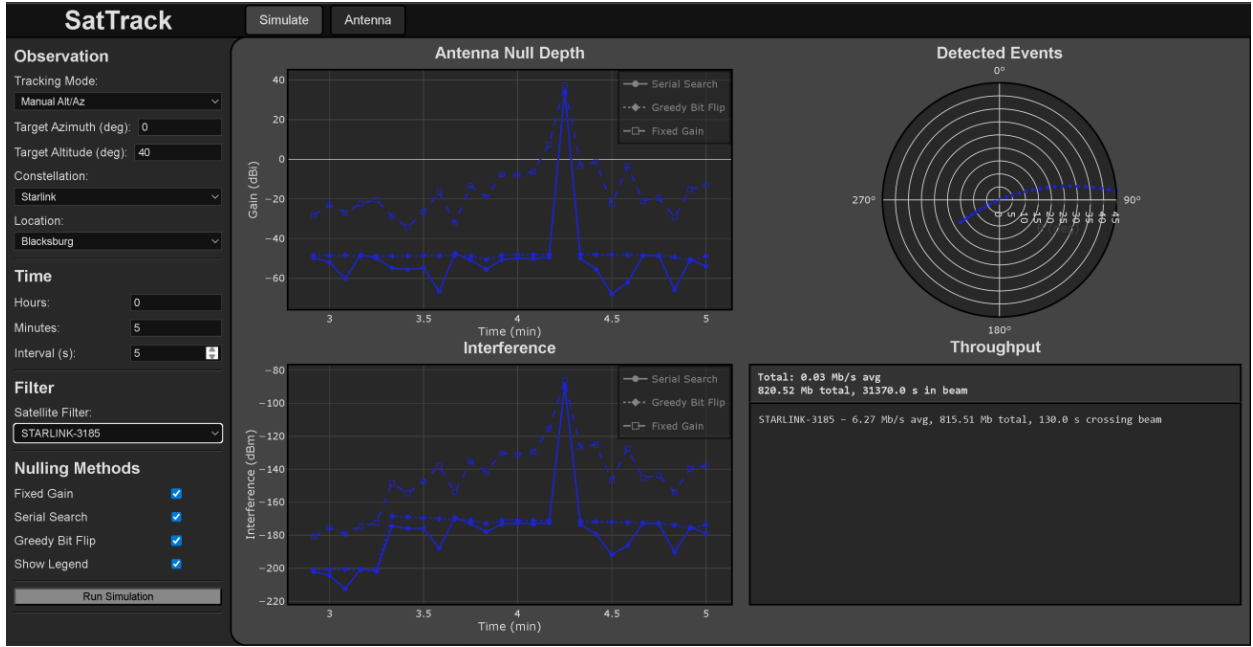

*Figure 17: The first instance of throughput using serial search interference mitigation.*

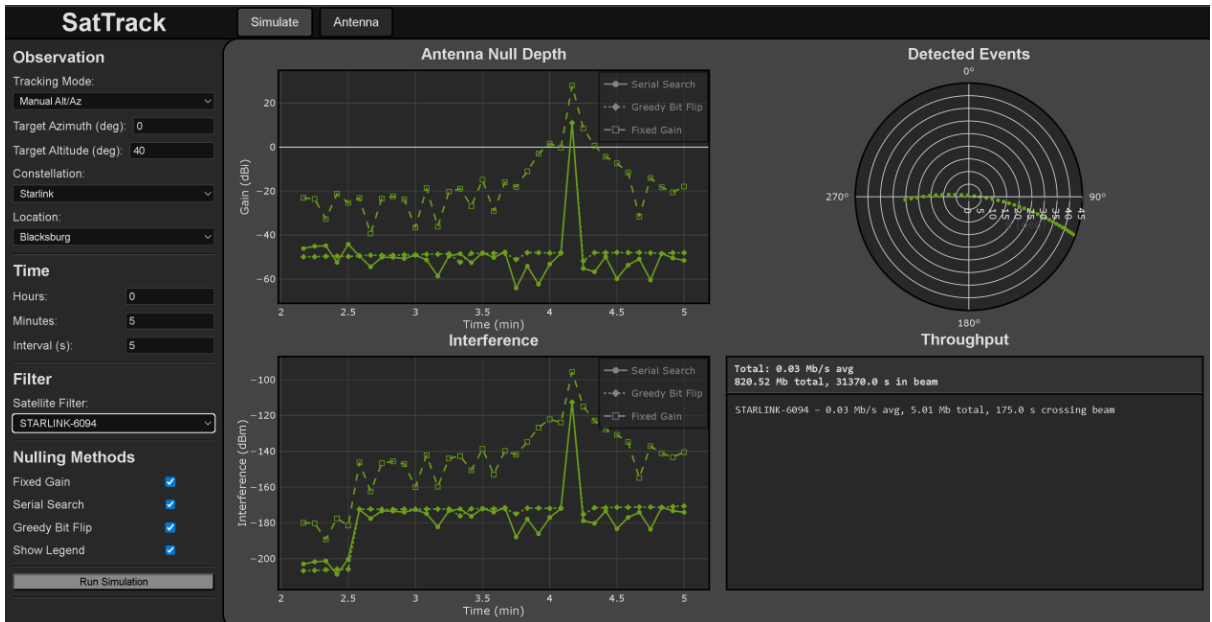

*Figure 18: The second instance of throughput using serial search interference mitigation.*

As shown in figure 17 & 18, a large spike is caused by STARLINK-3185 and STARLINK-6094 perfectly aligning with the main lobe.

## X. ADVANCED INTERFERENCE MITIGATION RESULTS

Simulations parameters:

- Antenna parameters:
    - Total Antenna Diameter: $2.5\ m$
    - Fixed Antenna Diameter: $2.222\ m$
    - Focal Length: $0.972\ m$
    - Transmit Frequency: $11.7\ GHz$
    - Feed Taper exponent: $1.14$
    - First rim-region width: $0.2\lambda$
    - Second rim-region width: $0.5\lambda$
    - Angular Sampling Step: $0.05°$
- Tracking parameters:
    - Observation From: Blacksburg
    - Satellite Constellation: Starlink
    - Observation Body: Polaris
    - Beamwidth: $5°, 15°, 45°$
    - Time Parameters: $0\ hours, 5\ minutes, 5\ seconds$ per sample

Weight alphabet convention:

- $M = 2$: $\{+1, -1\}$
    - Note that the serial search algorithm only uses binary weights.
- $M > 2$: points lie on the unit circle and avoid the axes.
    - Used by Simulated Annealing and MM
    - $W_M = \left\{ e^{j\left(\frac{\pi}{M} + \frac{2\pi m}{M}\right)} \mid m = 0, 1, \ldots, M-1 \right\}, (e.g., M = 4\ at\ 45°, 135°, 225°, 315°)$.

To run these tests yourself:

- Run in separate file "additional_tests.py"
    - Cd to main SatTrack Directory
    - Use command "python -m SatTrack.additional.additional_tests" in user terminal.

The notation for plots is the method (simulated annealing, MM) followed by the weight Mx (where x denotes weight alphabet used as described in the above section).

(Note: Runtime for 5 minutes of Starlink at 15° takes 45+ minutes)

### A. 5° from Boresight

The below results are for a 5-minute window of Starlink with 5° beamwidth having 23 detected satellites and 71 total events:

**Simulated Annealing:**

*Figure 19: Simulated Annealing M2 5° Boresight*

*Figure 20: Simulated Annealing M4 5° Boresight*

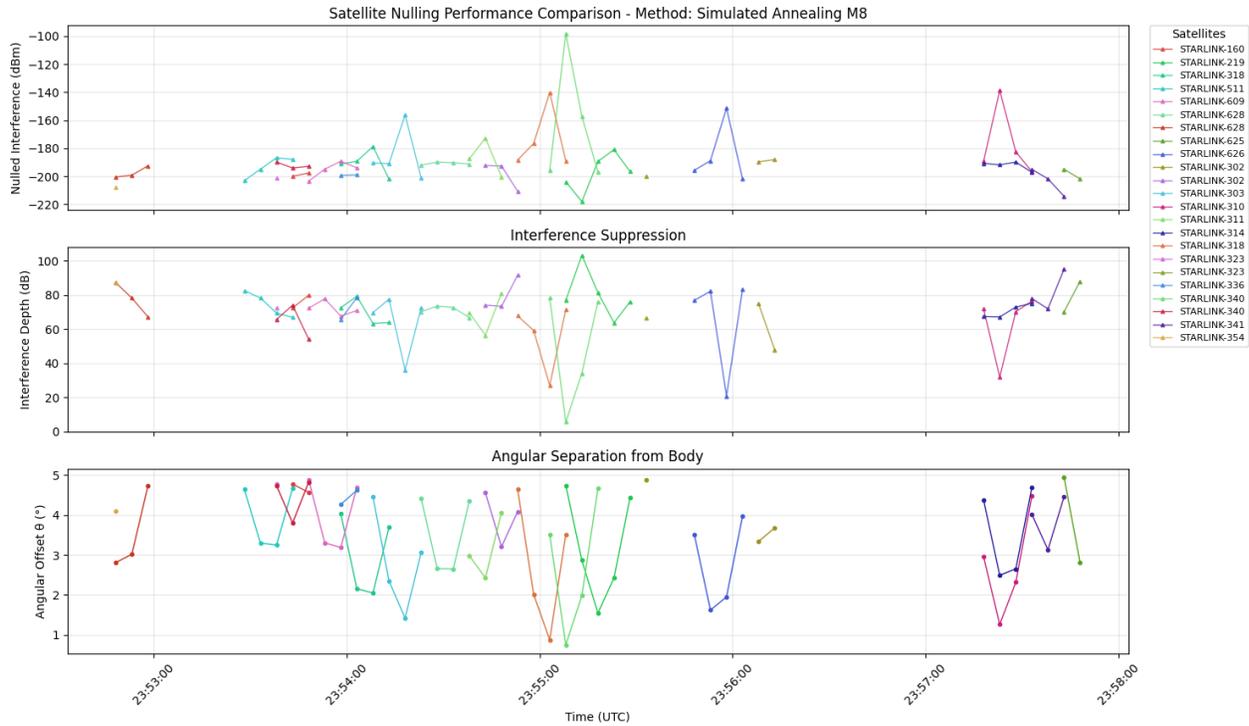

*Figure 21: Simulated Annealing M8 5° Boresight*

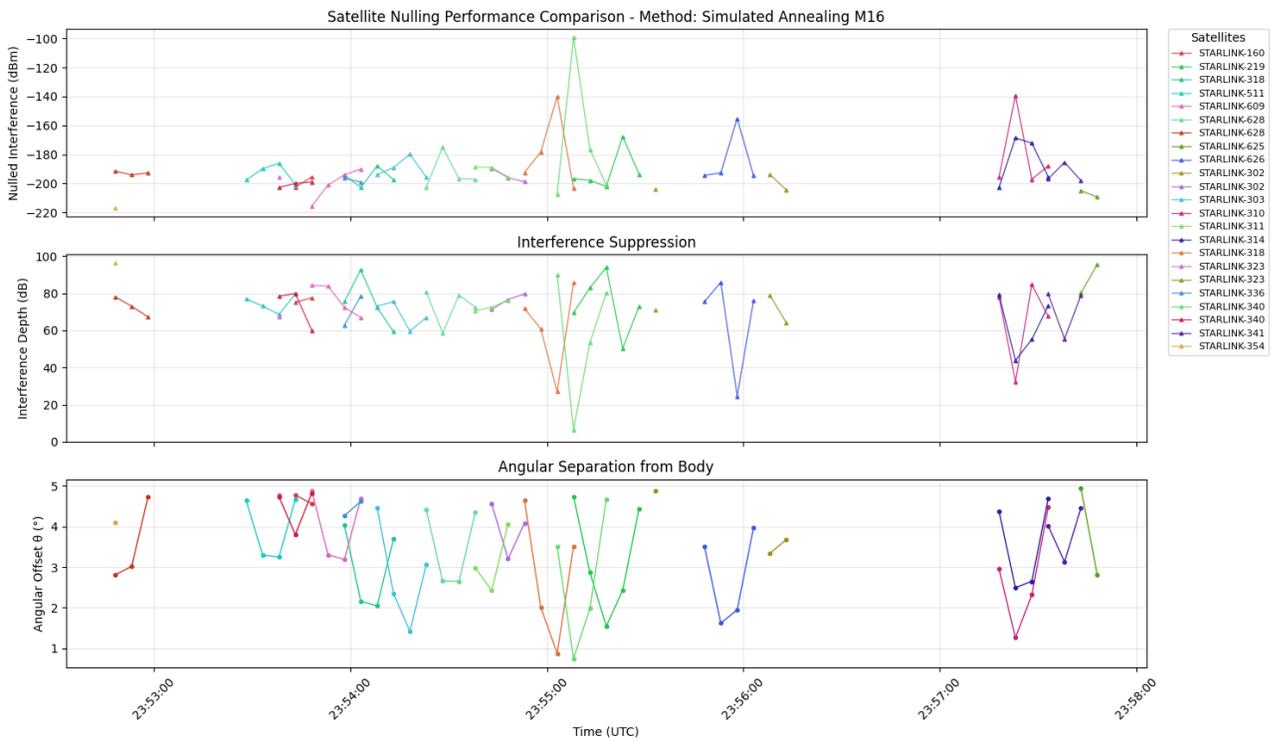

*Figure 22: Simulated Annealing M16 5° Boresight*

**MM:**

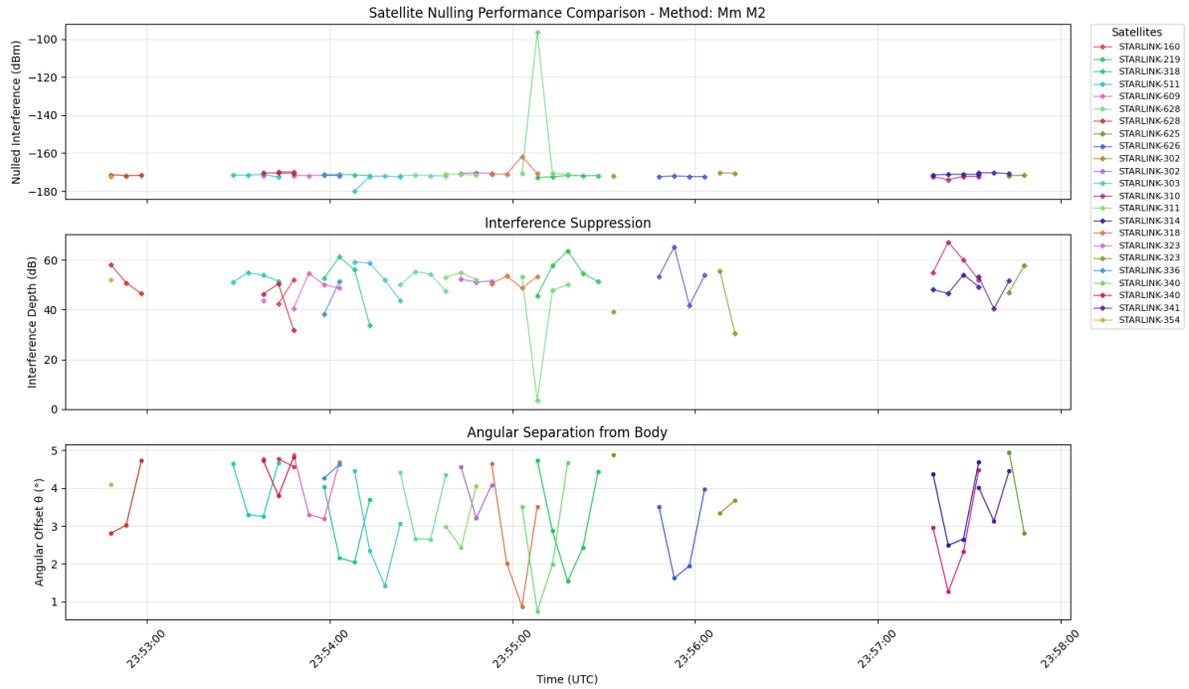

*Figure 23: MM M2 5° Boresight*

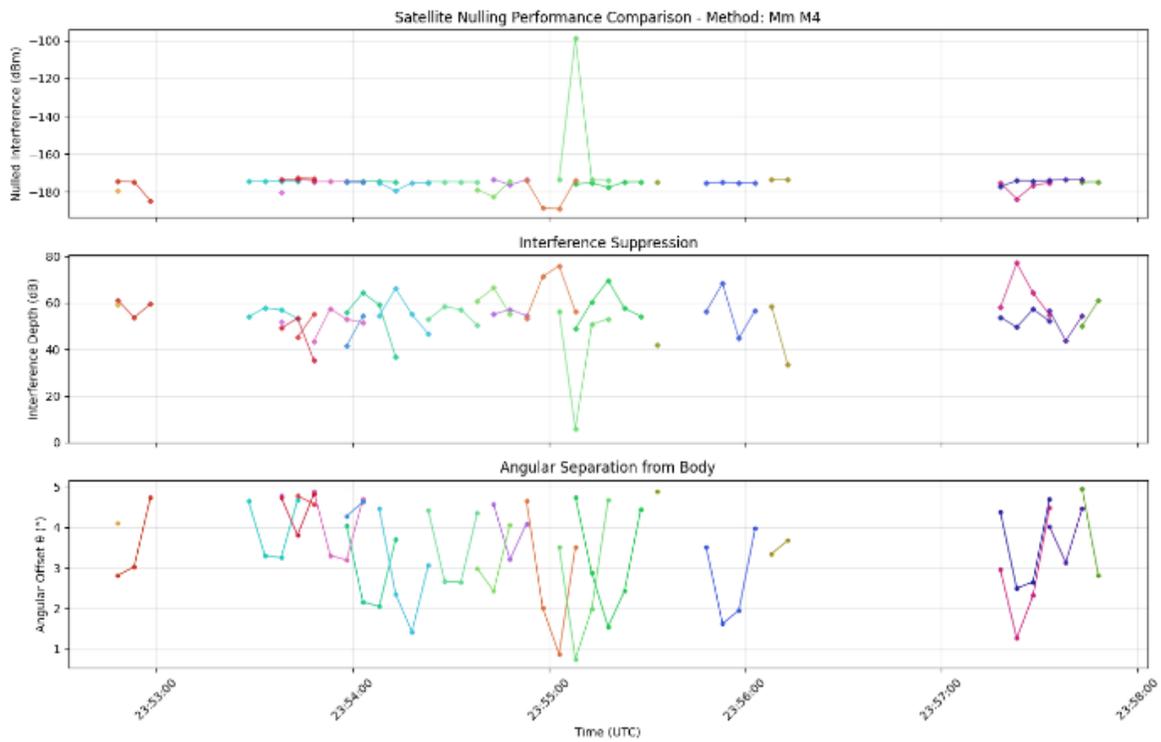

*Figure 24: MM M4 5° Boresight*

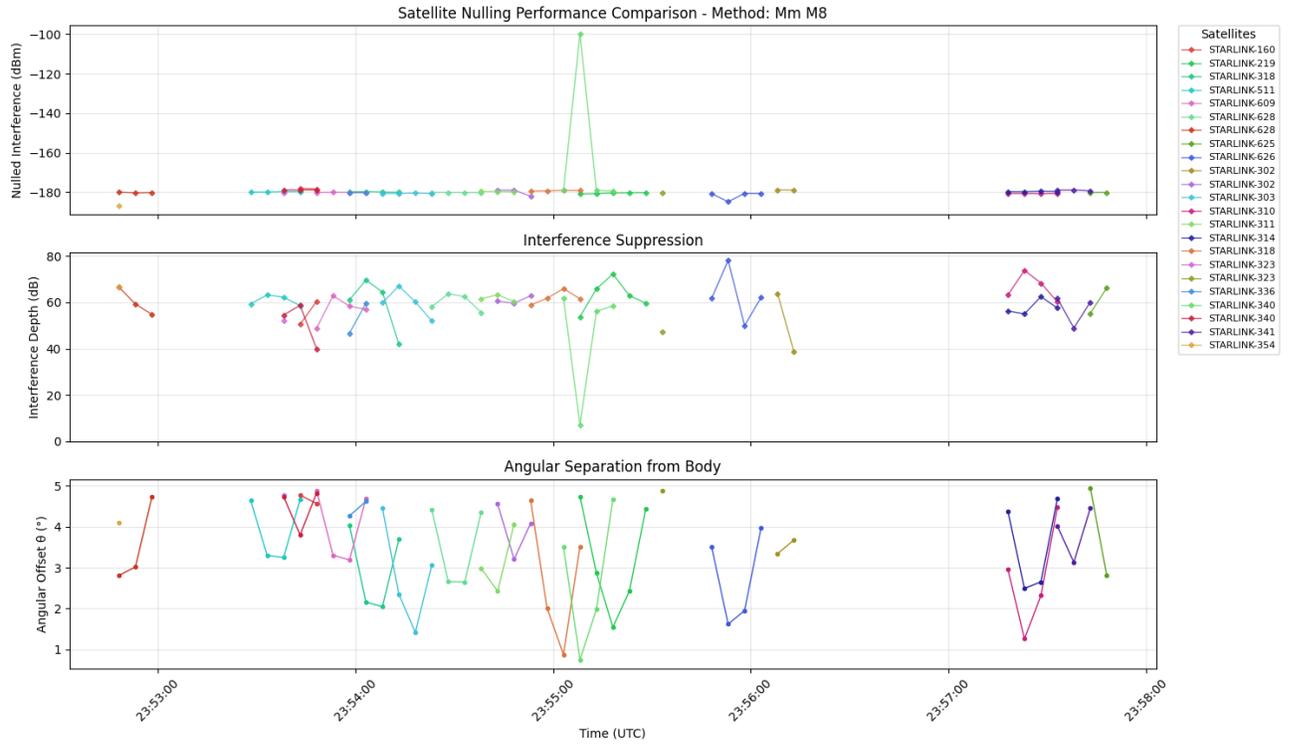

*Figure 25: MM M8 5° Boresight*

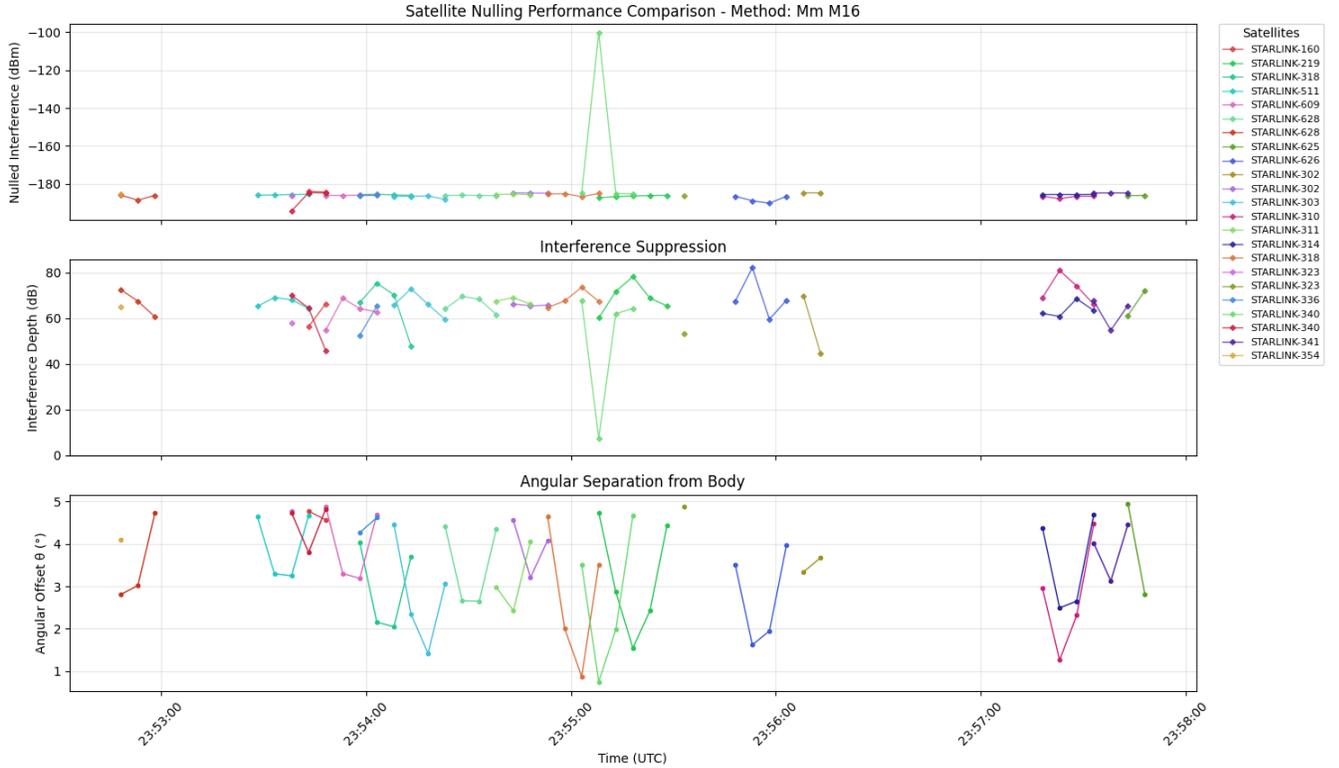

*Figure 26: MM M16 5° Boresight*

Methods Ranked:
1. Simulated Annealing M16:
    o   Average Suppression: 71.0 $dB$
    o   Lowest Received Power: $-216.9\ dBm$
2. Simulated Annealing M8:
    o   Average Suppression: 69.6 $dB$
    o   Lowest Received Power: $-218.1\ dBm$
3. Simulated Annealing M4:
    o   Average Suppression: 69.1 $dB$
    o   Lowest Received Power: $-220.3\ dBm$
4. Simulated Annealing M2:
    o   Average Suppression: 64.7 $dB$
    o   Lowest Received Power: $-200.4\ dBm$
5. MM M16:
    o   Average Suppression: 64.5 $dB$
    o   Lowest Received Power: $-194.3\ dBm$
6. MM M8:
    o   Average Suppression: 58.6 $dB$
    o   Lowest Received Power: $-186.9\ dBm$
7. SERIAL:
    o   Average Suppression: 55.1 $dB$
    o   Lowest Received Power: $-192.5\ dBm$
8. MM M4:
    o   Average Suppression: 54.4 $dB$
    o   Lowest Received Power: $-189.1\ dBm$
9. MM M2:
    o   Average Suppression: 50.2 $dB$
    o   Lowest Received Power: $-180.0\ dBm$

B.  $15°$   *from Boresight*

The below results are for a 5-minute window of Starlink with 15° beamwidth having 49 detected satellites and 437 total events:

**Simulated Annealing:**

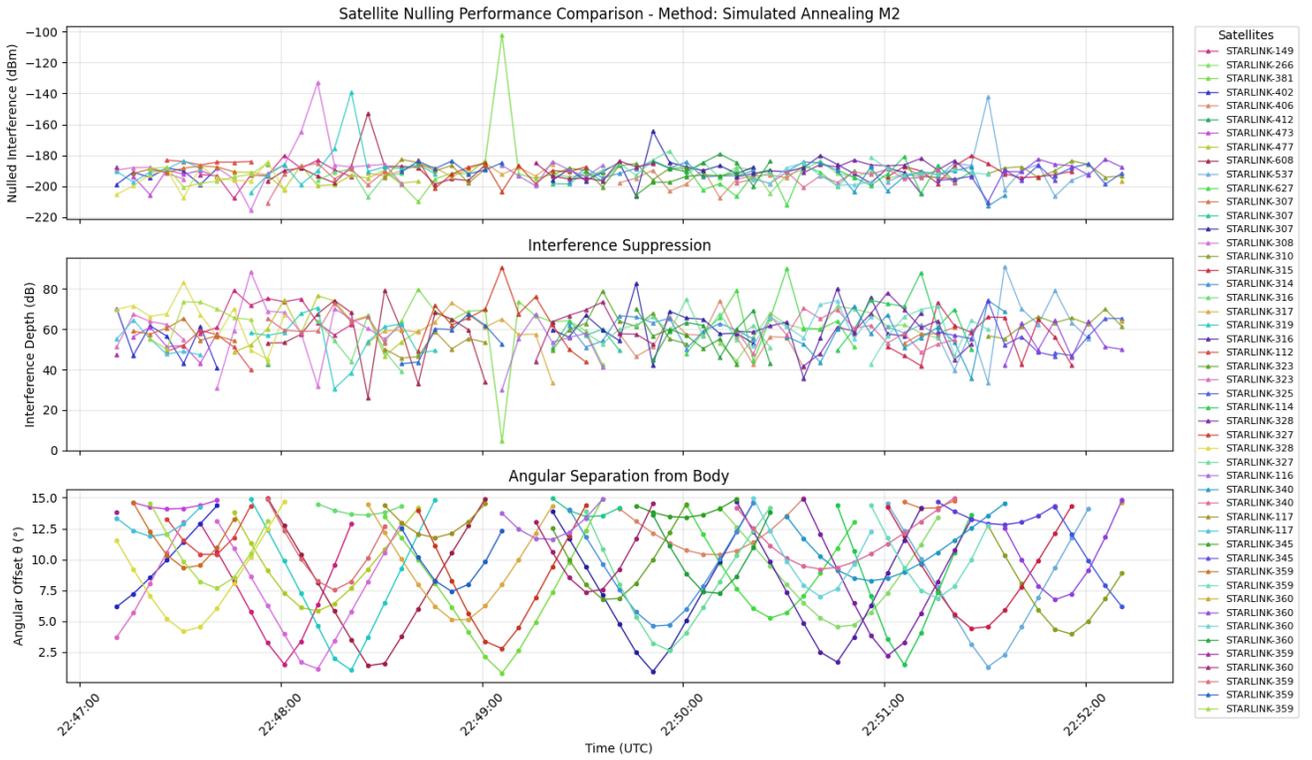

*Figure 27: Simulated Annealing M2* 15° *Boresight*

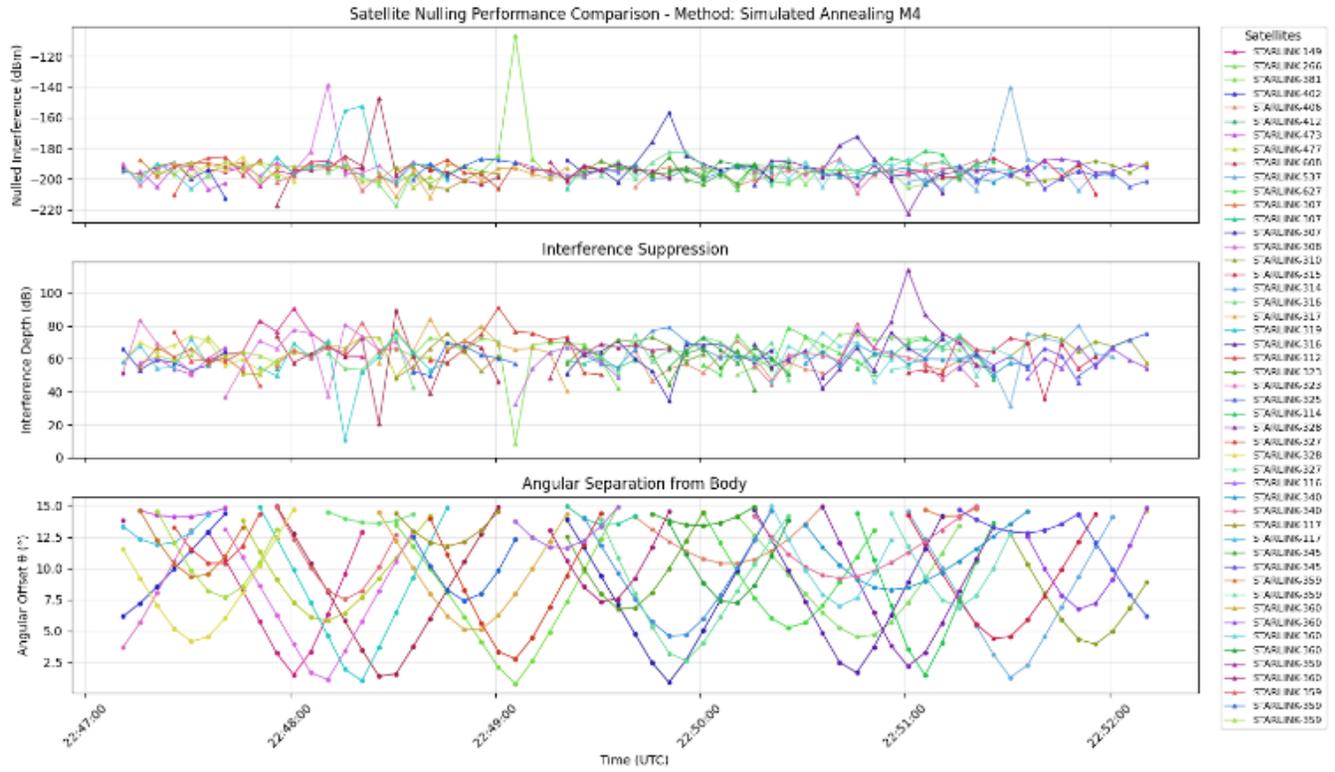

*Figure 28: Simulated Annealing M4* 15° *Boresight*

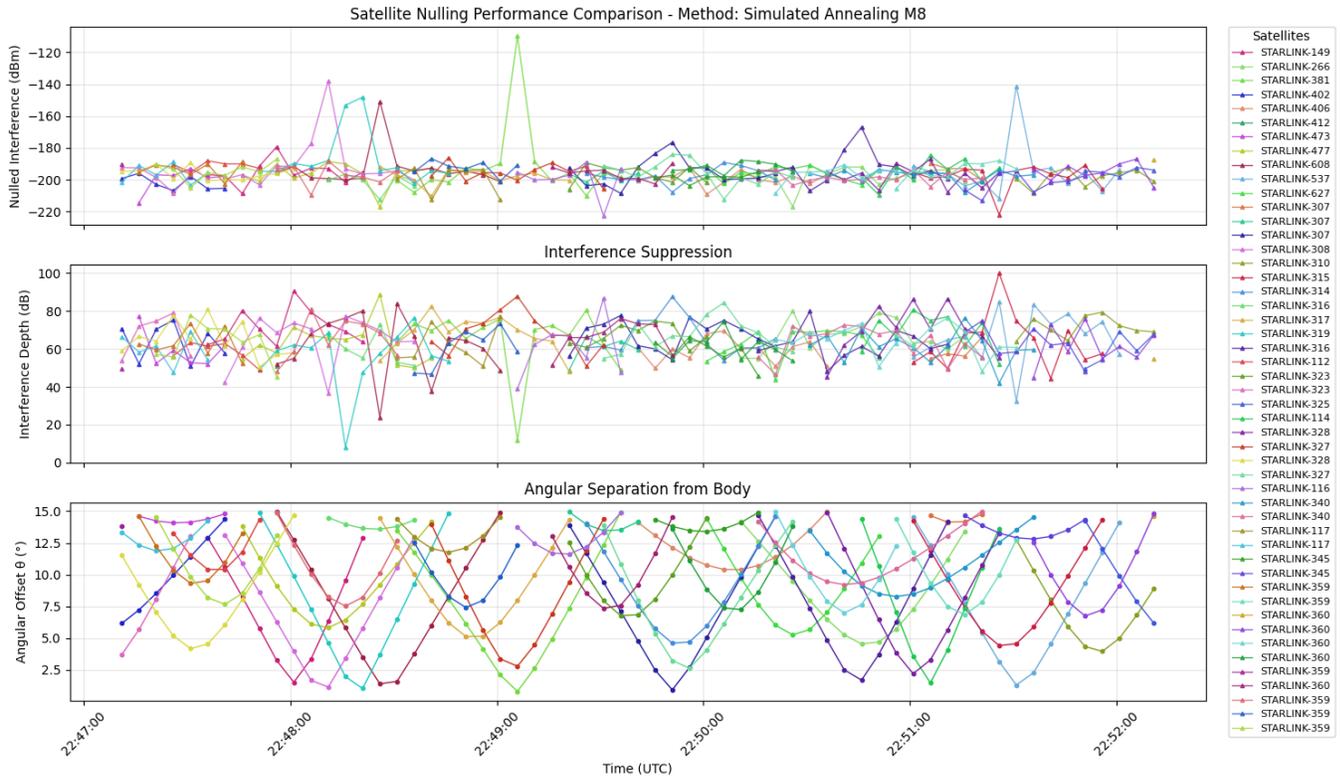

*Figure 29: Simulated Annealing M8* 15° *Boresight*

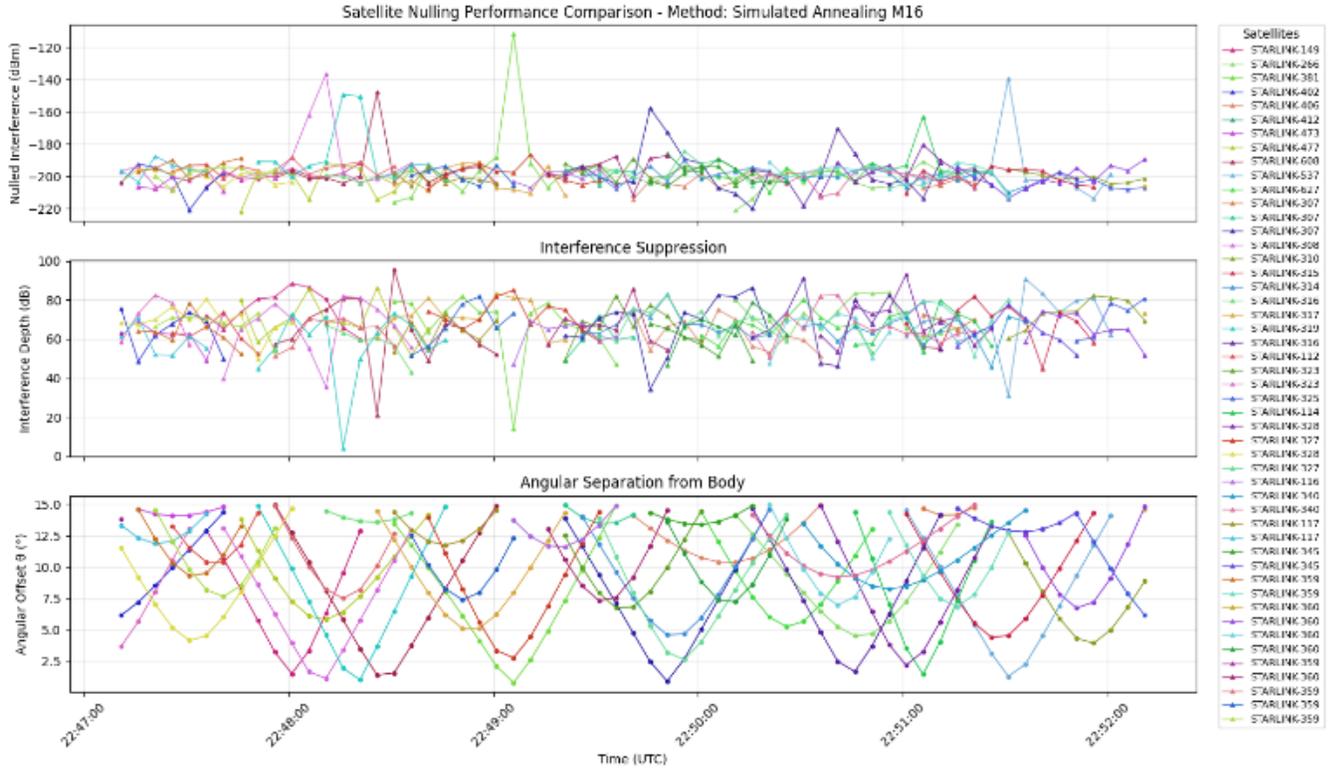

*Figure 30: Simulated Annealing M16 15° Boresight*

**MM:**

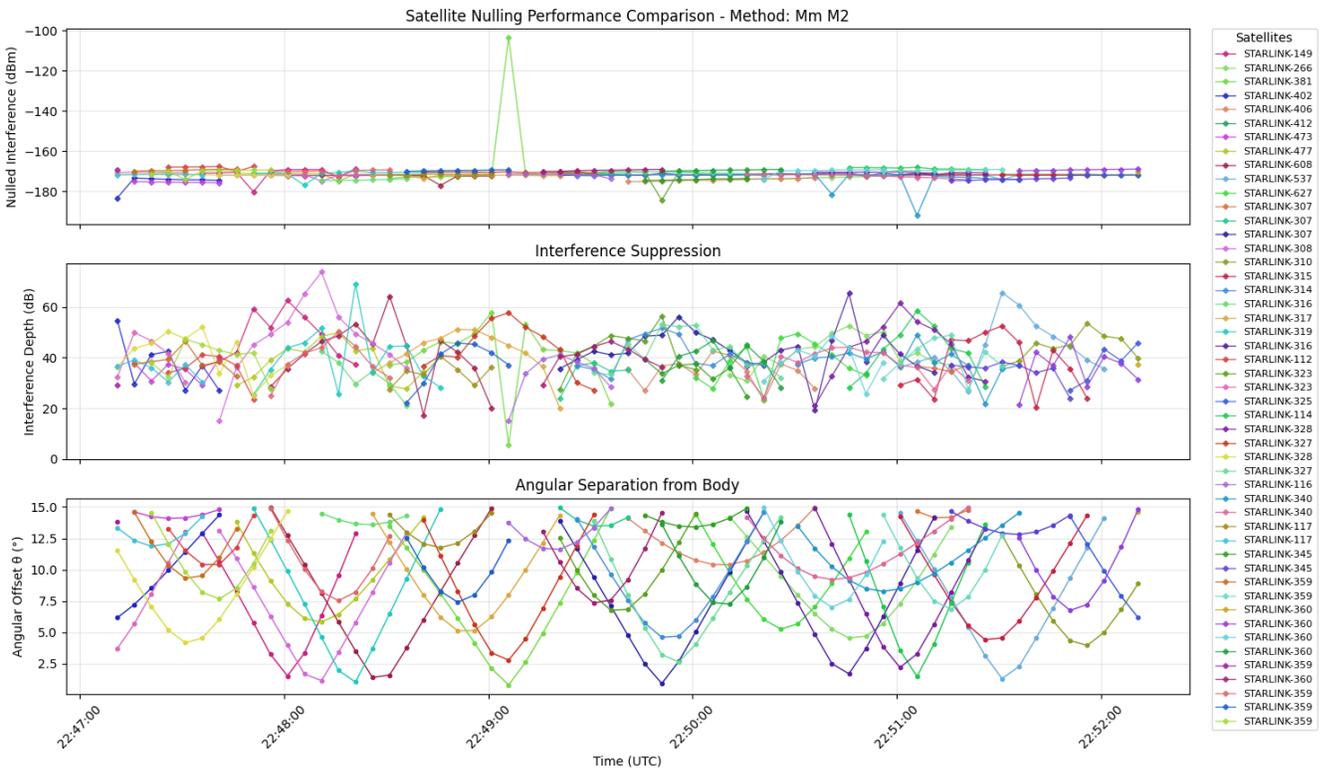

*Figure 31: MM M2 15° Boresight*

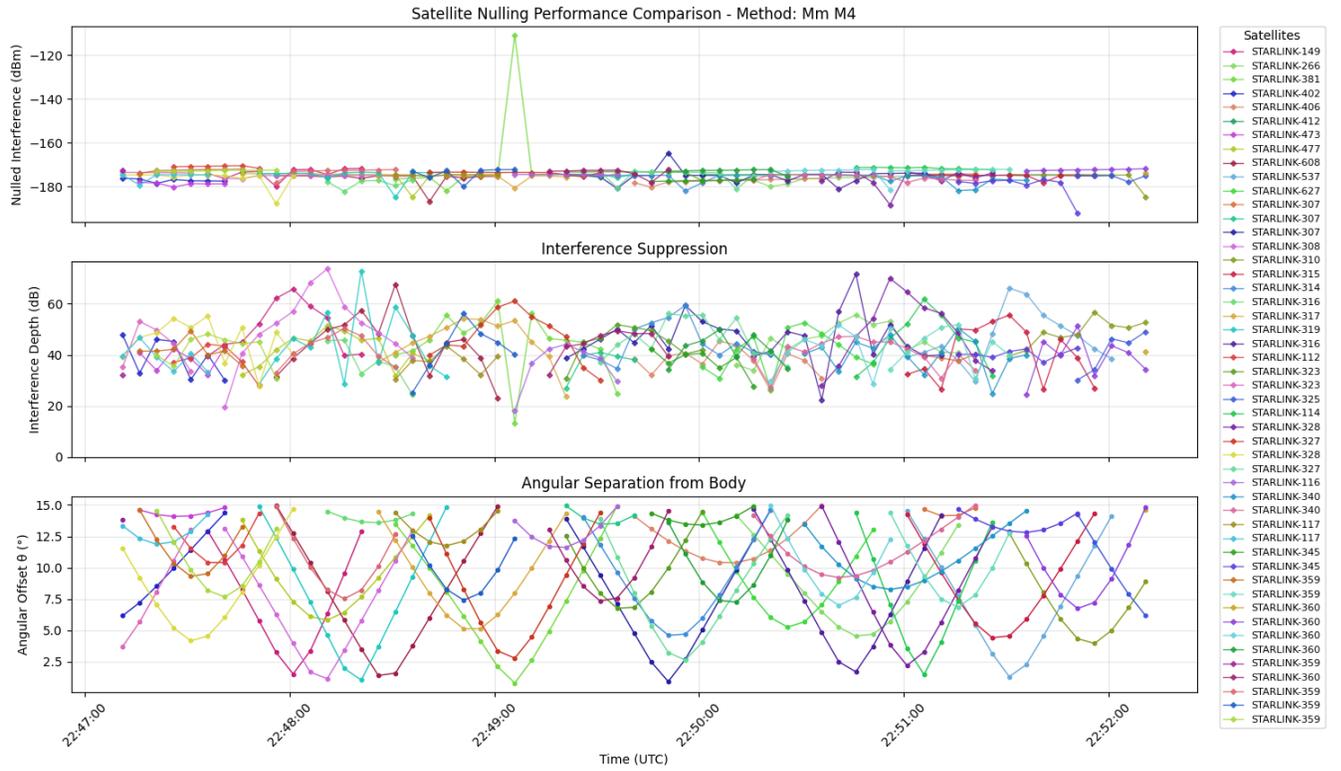

*Figure 32: MM M4 15° Boresight*

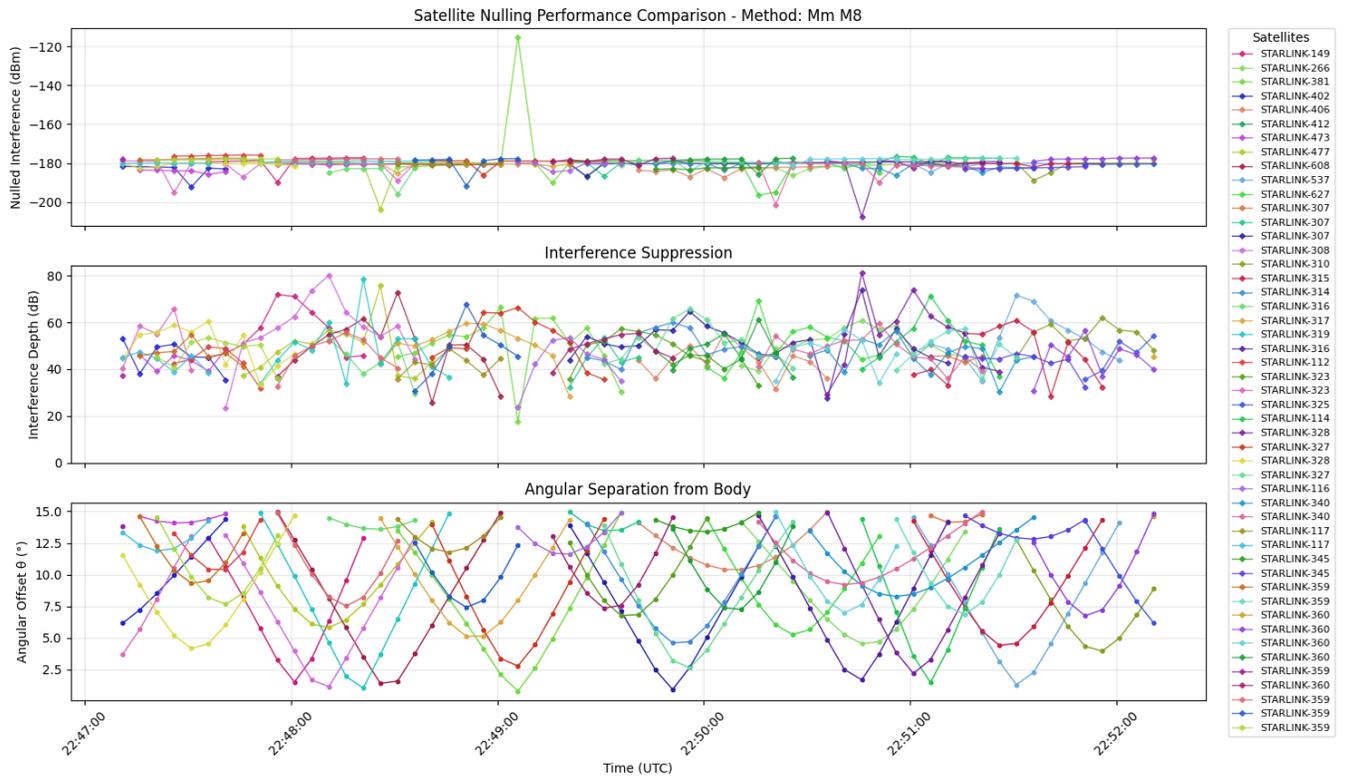

*Figure 33: MM M8 15° Boresight*

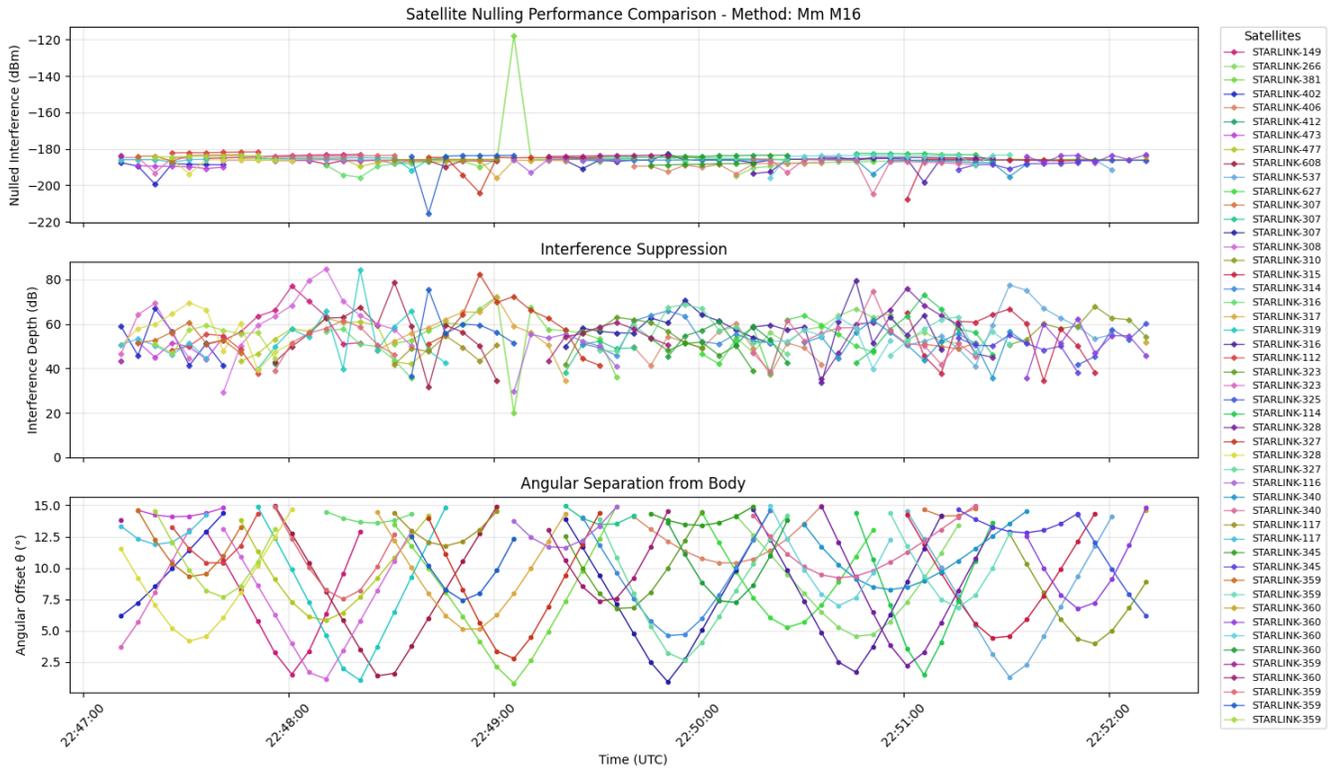

*Figure 34: MM M16 15° Boresight*

Methods Ranked:
1. Simulated Annealing M16:
    o Average Suppression: 66.4 *dB*
    o Lowest Received Power: −222.1 *dBm*
2. Simulated Annealing M8:
    o Average Suppression: 63.9 *dB*
    o Lowest Received Power: −222.7 *dBm*
3. Simulated Annealing M4:
    o Average Suppression: 62.1 *dB*
    o Lowest Received Power: −222.5 *dBm*
4. Simulated Annealing M2:
    o Average Suppression: 59.0 *dB*
    o Lowest Received Power: −215.5 *dBm*
5. MM M16:
    o Average Suppression: 54.3 *dB*
    o Lowest Received Power: −215.6 *dBm*
6. MM M8:
    o Average Suppression: 48.6 *dB*
    o Lowest Received Power: −207.6 *dBm*
7. SERIAL:
    o Average Suppression: 44.0 *dB*
    o Lowest Received Power: −199.6 *dBm*
8. MM M4:
    o Average Suppression: 43.0 *dB*
    o Lowest Received Power: −192.3 *dBm*
9. MM M2:
    o Average Suppression: 39.6 *dB*

- o  Lowest Received Power: $-192.0$ $dBm$

## C. 45°45º of Boresight

The below results are for a 5-minute window of Starlink with 45° beamwidth having 237 detected satellites and 6913 total events.

Note: The legend is inaccurate as it could not make a legend containing 237 different satellites

**Simulated Annealing:**

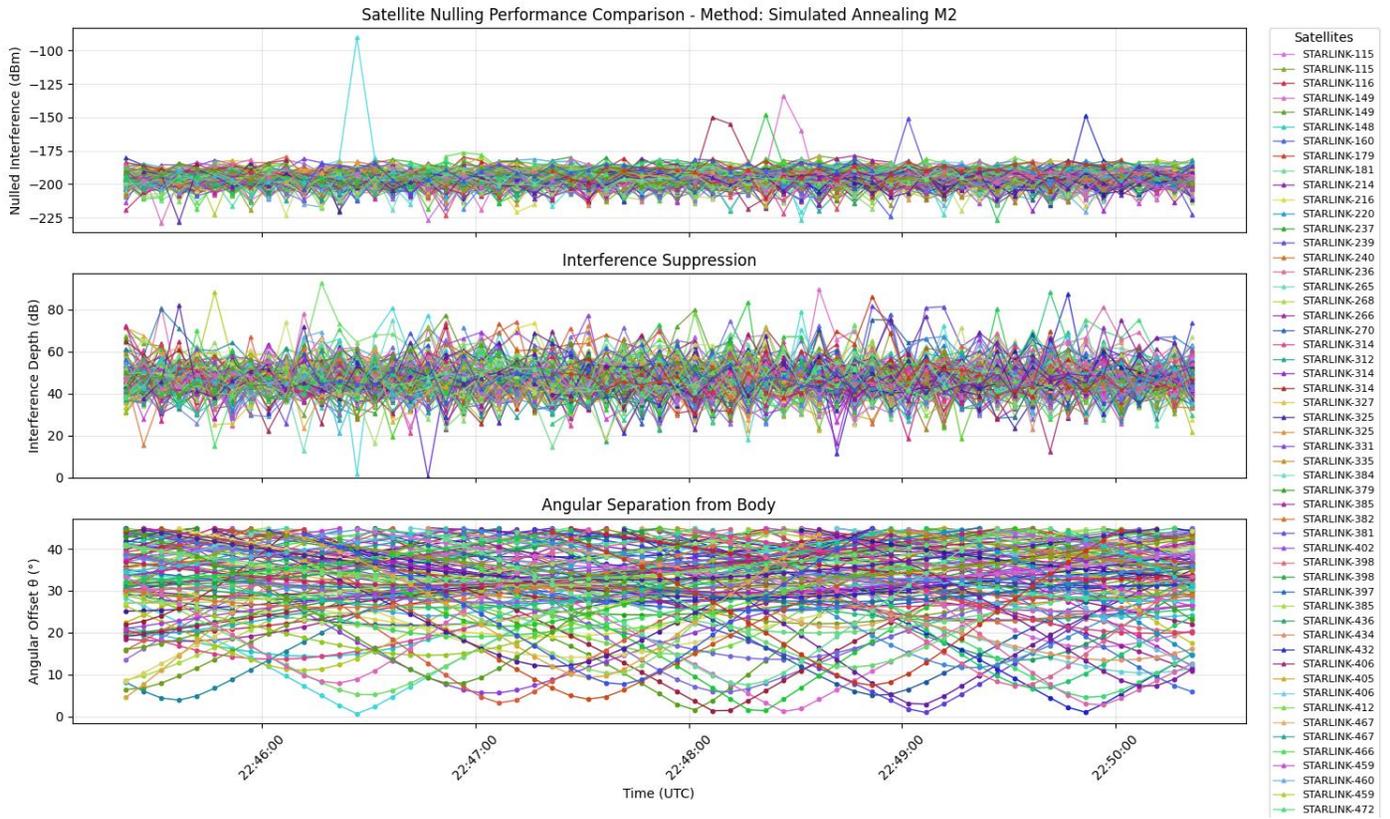

*Figure 35: Simulated Annealing M2 45° Boresight*

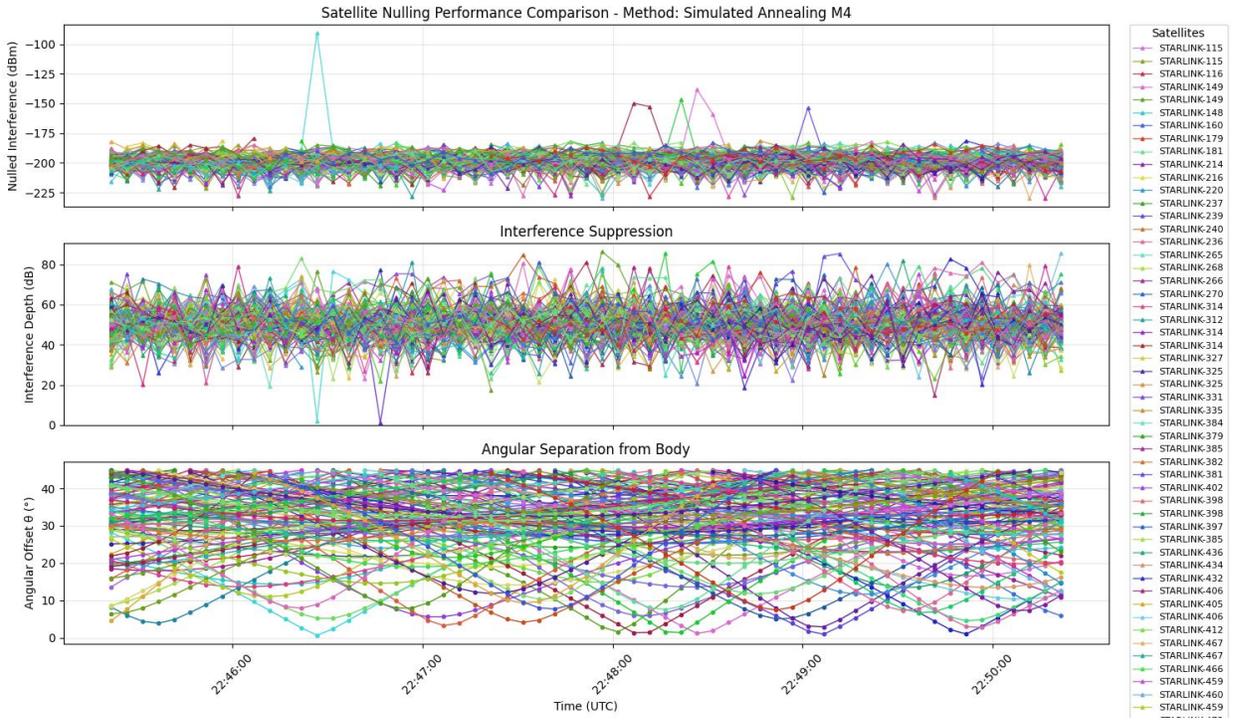

*Figure 36: Simulated Annealing M4 45° Boresight*

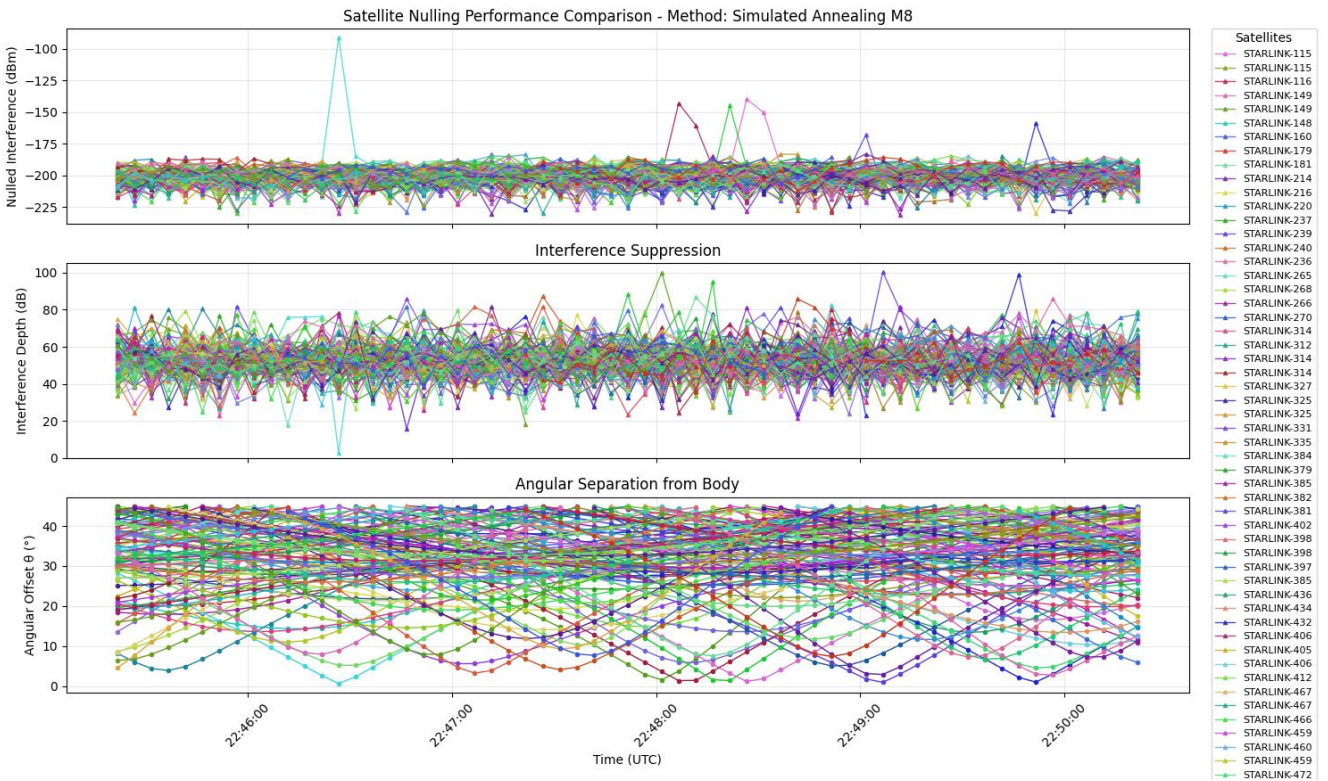

*Figure 37: Simulated Annealing M8 45° Boresight*

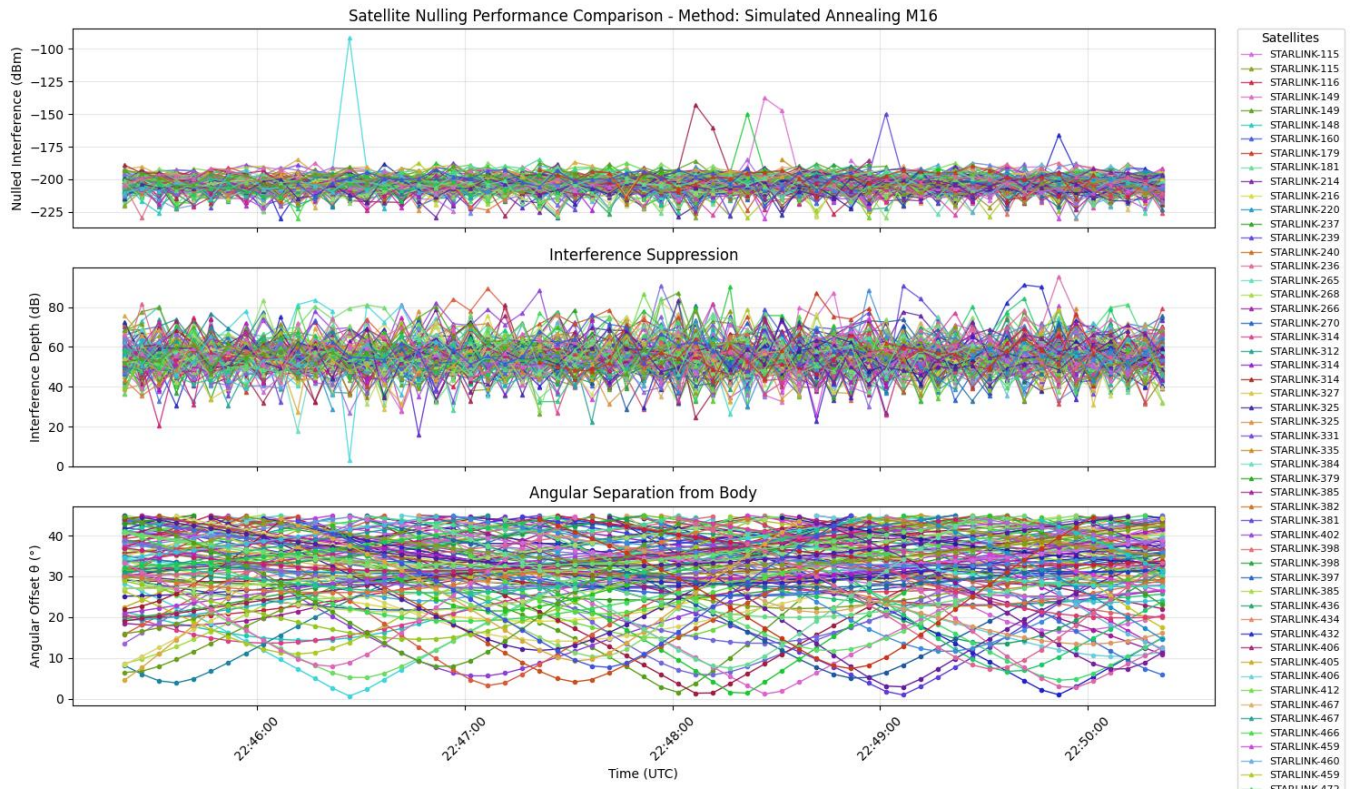

*Figure 38: Simulated Annealing M16 45° Boresight*

**MM:**

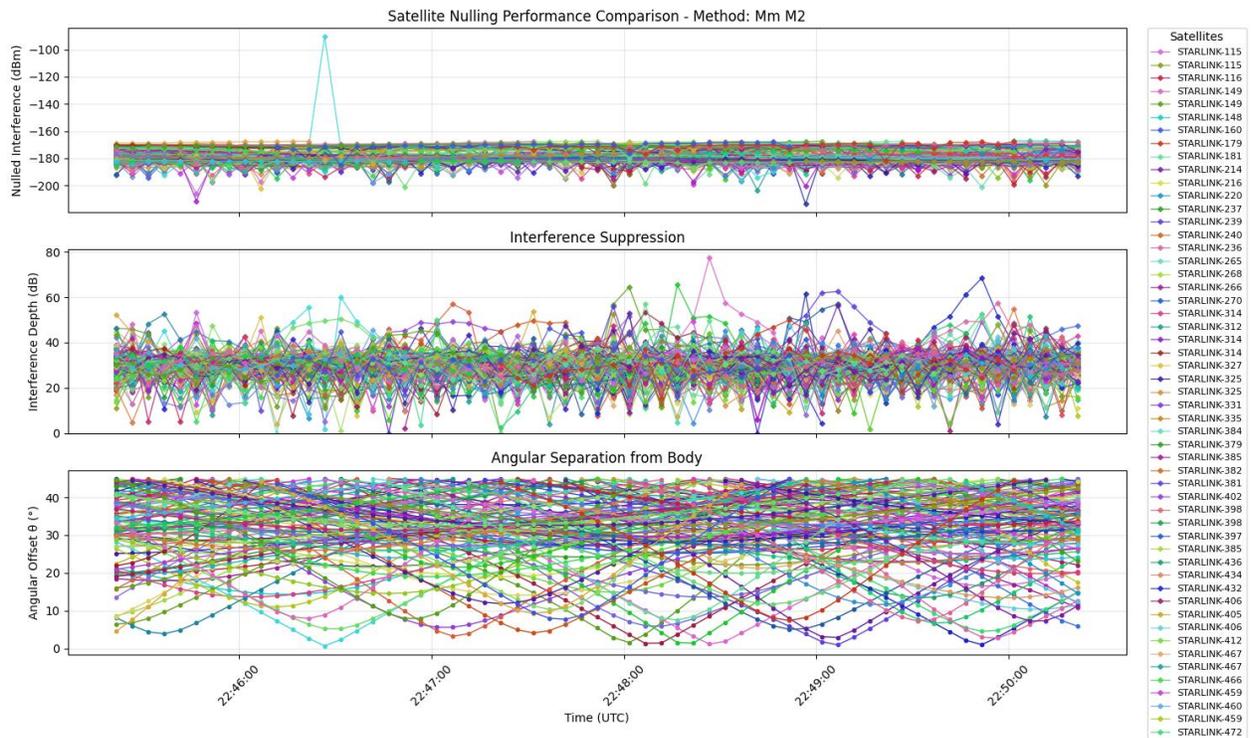

*Figure 39: MM M2 45° Boresight*

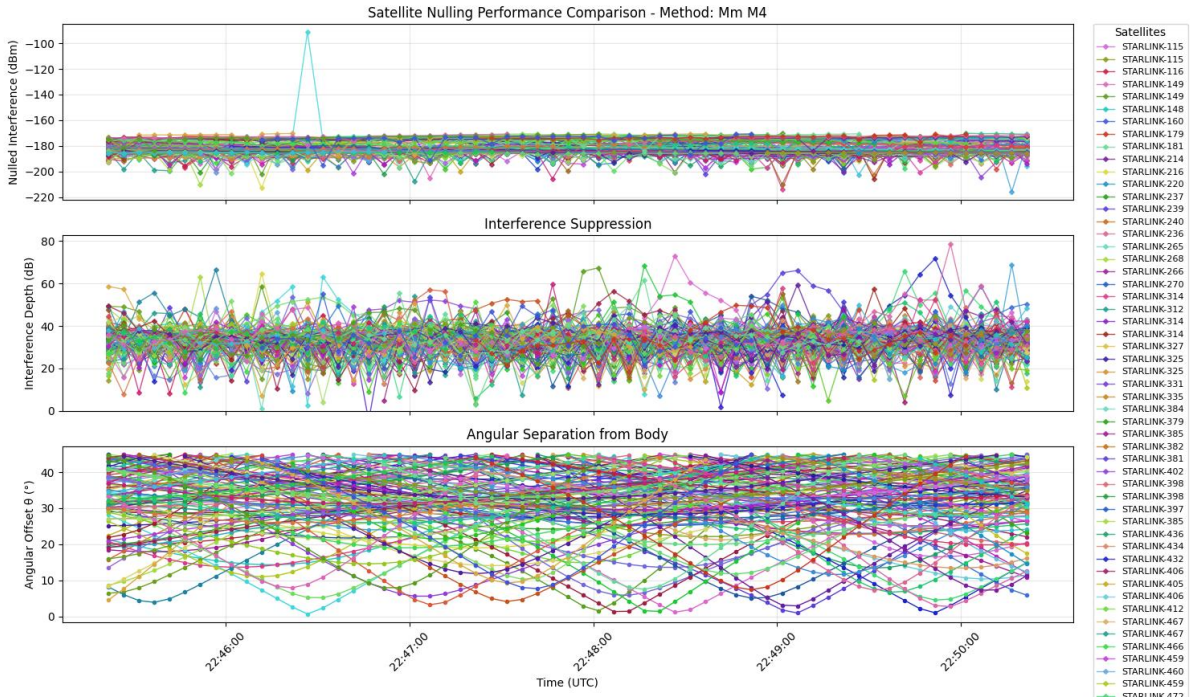

*Figure 40: MM M4 45° Boresight*

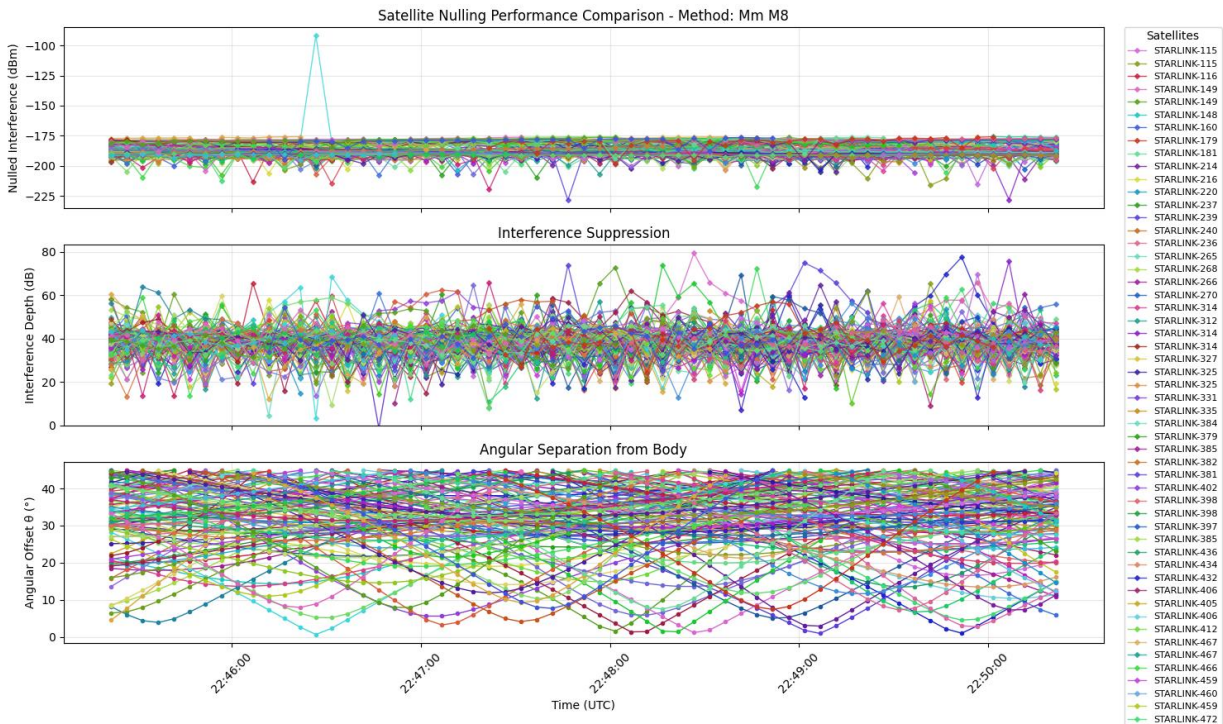

*Figure 41: MM M8 45° Boresight*

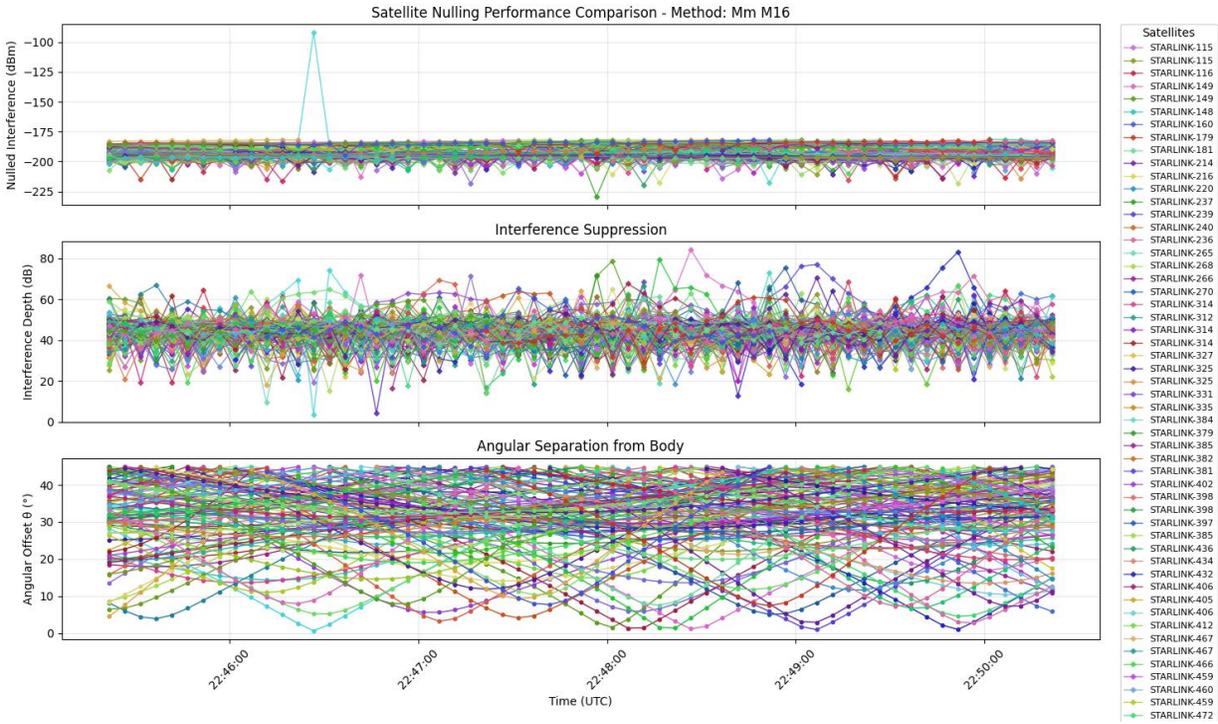

*Figure 42: MM M16 45° Boresight*

Methods Ranked:
1. Simulated Annealing M16:
    o   Average Suppression: 55.2 *dB*
    o   Lowest Received Power: −230.0 *dBm*
2. Simulated Annealing M8:
    o   Average Suppression: 52.5 *dB*
    o   Lowest Received Power: −231.1 *dBm*
3. Simulated Annealing M4:
    o   Average Suppression: 50.4 *dB*
    o   Lowest Received Power: −229.8 *dBm*
4. Simulated Annealing M2:
    o   Average Suppression: 47.1 *dB*
    o   Lowest Received Power: −229.2 *dBm*
5. MM M16:
    o   Average Suppression: 44.1 *dB*
    o   Lowest Received Power: −229.5 *dBm*
6. MM M8:
    o   Average Suppression: 38.4 *dB*
    o   Lowest Received Power: −228.4 *dBm*
7. MM M4:
    o   Average Suppression: 33.0 *dB*
    o   Lowest Received Power: −216.0 *dBm*
8. MM M2:
    o   Average Suppression: 29.9 *dB*
    o   Lowest Received Power: −213.6 *dBm*
9. SERIAL:
    o   Average Suppression: 29.9 *dB*
    o   Lowest Received Power: −195.1 *dBm*

XI.    TABLE SUMMARY OF RESULTS

| Considered Range of Angles | Method | Weight Alphabet | Avg. Interference Suppression (dB) | Lowest Interference Power (dBm) | Avg. Lost Throughput (Mb/s) | Total Lost Throughput (Mb) | Total Seconds in Beam |
|---|---|---|---|---|---|---|---|
| 5° | Turn Off Satellite | — | ∞ | −∞ | 160.34 | 28,059.6 | 230.0 |
| | None (Fixed) | — | 0 | −137.47 | 0 | 0 | 230.0 |
| | Serial | ±1 | 55.1 | −192.5 | ≈ 0 | ≈ 0 | 230.0 |
| | Simulated Annealing | M2 | 64.7 | −200.4 | ≈ 0 | ≈ 0 | 230.0 |
| | Simulated Annealing | M4 | 69.1 | −220.3 | ≈ 0 | ≈ 0 | 230.0 |
| | Simulated Annealing | M8 | 69.6 | −218.1 | ≈ 0 | ≈ 0 | 230.0 |
| | Simulated Annealing | M16 | 71.0 | −216.9 | ≈ 0 | ≈ 0 | 230.0 |
| | MM | M2 | 50.2 | −180.0 | ≈ 0 | ≈ 0 | 230.0 |
| | MM | M4 | 54.4 | −189.1 | ≈ 0 | ≈ 0 | 230.0 |
| | MM | M8 | 58.6 | −186.9 | ≈ 0 | ≈ 0 | 230.0 |
| | MM | M16 | 64.5 | −194.3 | ≈ 0 | ≈ 0 | 230.0 |
| 15° | Turn Off Satellite | — | ∞ | −∞ | 177.47 | 279,517.5 | 1750 |
| | None (Fixed) | — | 0 | −156.95 | ≈ 0 | ≈ 0 | 1750 |
| | Serial | ±1 | 44.0 | −199.6 | ≈ 0 | ≈ 0 | 1750 |
| | Simulated Annealing | M2 | 59.0 | −215.5 | ≈ 0 | ≈ 0 | 1750 |
| | Simulated Annealing | M4 | 62.1 | −222.5 | ≈ 0 | ≈ 0 | 1750 |
| | Simulated Annealing | M8 | 63.9 | −222.7 | ≈ 0 | ≈ 0 | 1750 |
| | Simulated Annealing | M16 | 66.4 | −222.1 | ≈ 0 | ≈ 0 | 1750 |
| | MM | M2 | 39.6 | −192.0 | ≈ 0 | ≈ 0 | 1750 |
| | MM | M4 | 43.0 | −192.3 | ≈ 0 | ≈ 0 | 1750 |
| | MM | M8 | 48.6 | −207.6 | ≈ 0 | ≈ 0 | 1750 |
| | MM | M16 | 54.3 | −215.6 | ≈ 0 | ≈ 0 | 1750 |
| 45° | Turn Off Satellite | — | ∞ | −∞ | 89.92 | 2,892,813 | 32170 |
| | None (Fixed) | — | 0 | −217.89 | ≈ 0 | ≈ 0 | 32170 |
| | Serial | ±1 | 29.9 | −195.1 | ≈ 0 | ≈ 0 | 32170 |
| | Simulated Annealing | M2 | 47.1 | −229.2 | ≈ 0 | ≈ 0 | 32170 |

| Considered Range of Angles | Method | Weight Alphabet | Avg. Interference Suppression (dB) | Lowest Interference Power (dBm) | Avg. Lost Throughput (Mb/s) | Total Lost Throughput (Mb) | Total Seconds in Beam |
|---|---|---|---|---|---|---|---|
| | Simulated Annealing | M4 | 50.4 | −229.8 | ≈ 0 | ≈ 0 | 32170 |
| | Simulated Annealing | M8 | 52.5 | −231.1 | ≈ 0 | ≈ 0 | 32170 |
| | Simulated Annealing | M16 | 55.2 | −230.0 | ≈ 0 | ≈ 0 | 32170 |
| | MM | M2 | 29.9 | −213.6 | ≈ 0 | ≈ 0 | 32170 |
| | MM | M4 | 33.0 | −216.0 | ≈ 0 | ≈ 0 | 32170 |
| | MM | M8 | 38.4 | −228.4 | ≈ 0 | ≈ 0 | 32170 |
| | MM | M16 | 44.1 | −229.5 | ≈ 0 | ≈ 0 | 32170 |

## XII. Discussion

Across all three considered ranges of angles, the reconfigurable rim based nulling techniques significantly reduced received interference power relative to the fixed reflector case while causing virtually no loss in satellite throughput. The strongest mitigation performance is observed when only attempting to null signals closer to boresight, with approximately 55 $dB$ average interference reduction when nulling within 5° of boresight. As the angle range increases, the average null depth decreases modestly due to increased event density.

This correlation is also evident in the advanced methods section, which incorporates additional mitigation techniques. Overall, the results from both the simple and advanced interference mitigation algorithms show that, regardless of how effective the mitigation is, while satellites crossing directly through the main lobe cannot be mitigated, those events are rare (less than 0.03% chance of occurring).

Throughput results show that although the average throughput per satellite decreases at large boresight angles, the total accumulated throughput increases substantially due to the much longer aggregate dwell time of satellites within the beam. This highlights the trade-off between instantaneous link quality and total data volume as a function of observation geometry. Furthermore, using serial search to mitigate interference results in a theoretical loss in received throughput of ~0 $Mb$ from Starlink satellites.

## XIII. Contributions

Sections I-IX were written by Justin Santana, with all code relating to the GUI, antenna simulation, satellite tracking, SatTrack API, and serial search/greedy bit-flip optimization also written by Justin Santana. Developer credit is given at the top of each file in the repository. Section X, along with code relating to advanced interference mitigation efforts (simulated annealing / MM) was written by Luke Heller. Dr. Mike Buehrer supervised this work.

The GitHub repository containing the code can be found here: https://github.com/Neo630/SatTrack.git.

**This work was supported in part by the U.S. National Science Foundation under Grant AST-2128506.**

## References

[1] S. Ellingson and R. Sengupta, "Sidelobe Modification for Reflector Antennas by Electronically Reconfigurable Rim Scattering," in *IEEE Antennas and Wireless Propagation Letters*, vol. 20, no. 6, pp. 1083-1087, June 2021, doi: 10.1109/LAWP.2021.3072536.

[2] R. M. Buehrer and S. W. Ellingson, "Pattern Control for Reflector Antennas Using Electronically-Reconfigurable Rim Scattering," 2022 IEEE International Symposium on Antennas and Propagation and USNC-URSI Radio Science Meeting (AP-S/URSI), Denver, CO, USA, 2022, pp. 577-578, doi: 10.1109/AP-S/USNC-URSI47032.2022.9886543.

[3] R. M. Buehrer, W. W. Howard, and S. Ellingson, "Open and closed-loop weight selection for pattern control of paraboloidal reflector antennas with reconfigurable rim scattering," *arXiv preprint* arXiv:2308.16339, Aug. 2023, doi: 10.48550/arXiv.2308.16339.

[4] X. Li, R. M. Buehrer and S. W. Ellingson, "An Improved Weight Selection Algorithm for Interference Mitigation in Paraboloidal Reflector Antennas with Reconfigurable Rim Scattering," MILCOM 2024 - 2024 IEEE Military Communications Conference (MILCOM), Washington, DC, USA, 2024, pp. 1-7, doi: 10.1109/MILCOM61039.2024.10773626.

[5] "SPACEX NON-GEOSTATIONARY SATELLITE SYSTEM ATTACHMENT A TECHNICAL INFORMATION TO SUPPLEMENT SCHEDULE S A.1 SCOPE AND PURPOSE," 2018. Available: https://fcc.report/IBFS/SAT-MOD-20181108-00083/1569860.pdf

[6] "Application Form SES-LIC-INTR2021-02141," *Fcc.report*, 2021. https://fcc.report/IBFS/SES-LIC-INTR2021-02141/9029804